\documentclass[letter,12pt]{article}
\usepackage[T1]{fontenc}
\usepackage[dvips]{graphicx}
\graphicspath{{images/}}
\usepackage{verbatim}
\usepackage{amsmath,amsthm}
\usepackage{xspace}
\usepackage{pifont}
\usepackage{graphicx}
\usepackage{amssymb}
\usepackage{epic, eepic}
\usepackage{dsfont}
\usepackage{amssymb}
\usepackage{makeidx}
\usepackage{mathrsfs}
\usepackage{exscale}
\usepackage{color} 
\usepackage{overpic} 
\usepackage{bm}
\usepackage{bbm}
\usepackage{booktabs} % To thicken table lines
\usepackage{color, colortbl}
\usepackage{subcaption}

\definecolor{Gray}{gray}{0.9}

\usepackage{amsmath,afterpage}
\usepackage{epsf}
\usepackage{graphics,color}

\def\0{\mathbf{0}}

\def\calP{{\cal P}}
\def\calT{{\cal T}}

\def\lam{\lambda}
\def\rr{\rightarrow}

\def \< {\langle}
\def \> {\rangle}

\def\beqa{\begin{eqnarray}}
\def\eeqa{\end{eqnarray}}
\def\beqas{\begin{eqnarray*}}
\def\eeqas{\end{eqnarray*}}

\newtheorem{theorem}{Theorem}[section]

\newtheorem{proposition}[theorem]{Proposition}

\newtheorem{corollary}[theorem]{Corollary}

\newtheorem{remark}[theorem]{Remark}

\newtheorem{definition}[theorem]{Definition}

\numberwithin{equation}{section}
\newcommand{\hatd}[1]{{}}

\newcommand{\bd}{\begin{displaymath}}
\newcommand{\ed}{\end{displaymath}}
\newcommand{\be}{\begin{equation}}
\newcommand{\ee}{\end{equation}}
\newcommand{\bq}{\begin{eqnarray}}
\newcommand{\eq}{\end{eqnarray}}
\newcommand{\bn}{\begin{eqnarray*}}
\newcommand{\en}{\end{eqnarray*}}
\newcommand{\dl}{\delta}
\newcommand{\re}{\mathds{R}}

\def\wt{\widetilde}

\def\one{\mathds{1}}

\usepackage{authblk}

\title{Incorporating Signals  into Optimal Trading}
\title{Incorporating Signals  into Optimal Trading}
\author[1,2]{Charles-Albert Lehalle }
\author[2,3]{Eyal Neuman \thanks{http://eyaln13.wixsite.com/eyal-neuman}}
\affil[1]{Capital Fund Management, Paris }
\affil[2]{CFM-Imperial College Institute, London}
\affil[3]{Department of Mathematics, Imperial College London }

\begin{document}

 \vspace{-0.5cm}
\maketitle

\begin{abstract}
  Optimal trading is a recent field of research which was initiated by Almgren, Chriss, Bertsimas and Lo in the late 90's.
Its main application is slicing large trading orders, in the interest of minimizing trading costs and potential perturbations of price dynamics due to liquidity shocks.  The initial optimization frameworks were based on mean-variance minimization for the trading costs. In the past 15 years, finer modelling of price dynamics, more realistic
control variables and different cost functionals were developed. The inclusion of signals (i.e. short term predictors of price dynamics) in optimal trading is a recent development and it is also the subject of this work. 

We incorporate a Markovian signal in the optimal trading framework which was initially proposed by
 Gatheral, Schied, and Slynko \cite{GSS} and provide results on the existence and uniqueness of an optimal trading strategy. 
Moreover, we derive an explicit singular optimal strategy for the special case of an Ornstein-Uhlenbeck signal and an exponentially decaying transient market impact.
The combination of a mean-reverting signal along with a market impact decay is of special interest, since they affect the short term price variations in opposite directions.

Later, we show that in the asymptotic limit were the transient market impact becomes instantaneous, the optimal strategy becomes continuous. This result is compatible with the optimal trading framework which was proposed by Cartea and Jaimungal \cite{CarteaJaimungal1}.

In order to support our models, we analyse nine months of tick by tick data on 13 European stocks from the NASDAQ OMX exchange. We show that orderbook imbalance is a predictor of the future price move and it has some mean-reverting properties. From this data we show that market participants, especially high frequency traders, use this signal in their trading strategies. 
\end{abstract}

\section{Introduction}

The financial crisis of 2008-2009 raised concerns about the inventories kept by intermediaries. % (mainly investment banks). 
Regulators and policy makers took advantage of two main regulatory changes (Reg NMS in the US and MiFID in Europe) which were followed by the creation of worldwide trade repositories. They also enforced more transparency on the transactions and hence on market participants positions, which pushed the trading processes toward electronic platforms \cite{citeulike:12047995}.
Simultaneously, consumers and producers of financial products asked for less complexity and more transparency.

This tremendous pressure on the business habits of the financial system, shifted it from a customized and high margins industry, in which intermediaries could keep large (and potentially risky) inventories, to a mass market industry where logistics have a central role. As a result, investment banks nowadays unwind their risks as fast as possible.
%: first to satisfy regulatory capital constraints, and second because the demand of transparency %prevents % them %to make large margins while protecting the 
%too keep secret their inventory imbalance.
%would expose their potential inventory to everyone's eyes.
In the context of small margins and high velocity of position changes, trading costs %(and thus the optimization of the trading process) 
are of paramount importance. A major factor of the trading costs is the market impact: the faster the trading rate, the more the buying or selling pressure will move the price in a detrimental way. %But on the other hand, the slower the trading rate, the further the transaction price from the decision price.

Academic efforts to reduce the transaction costs of large trades started with the seminal papers of Almgren and Chriss \cite{OPTEXECAC00} and Bertsimas and Lo \cite{BLA98}. Both models deal with %identified the need for optimizing 
the trading process of one large market participant (for instance an asset manager or a bank) who would like to buy or sell a large amount of shares or contracts during a specified duration. %In practice, this reference duration is typically a trading day.
The cost minimization problem turned out to be quite involved, due to multiple constraints on the trading strategies.
On one hand, the %now well known 
market impact (%i.e. the price move due to the buying or selling pressure of the large order moves the price in a detrimental direction, 
see \cite{citeulike:13497373} and references therein) demands to trade slowly, or at least at a pace which takes into account the available liquidity.
On the other hand, traders have an incentive to trade rapidly, because they do not want to carry the risk of an adverse price move far away from their decision price.

%Optimal trading is about dealing with market impact to decrease transaction costs, taking into account that trading too slow would expose a trader to adversarial price changes with respect of its decision price.
The importance of optimal trading in the industry generated a lot of variations for the initial mean-variance minimization of the trading costs (see \cite{citeulike:12047995,cartea15book,olivier16book} for details). In this paper, we consider the mean-variance minimization problem in the context of stochastic control (see e.g. \cite{citeulike:5282787}, \cite{citeulike:5797837}). In this approach some more realistic control variables which are related to order book dynamics and specific stochastic processes for the underlying price can be used (see \cite{GLFT} and \cite{citeulike:13675369} for related work). 

In this paper we address the question of how to incorporate \emph{signals}, which are predicting short term price moves, into optimal trading problems.
Usually optimal execution problems focus on the tradeoff between market impact and market risk. However, in practice many traders and trading algorithms use short term price predictors. 
Most of such documented predictors relate to orderbook dynamics \cite{citeulike:12820703}.
They can be divided into two categories: signals which are based on liquidity consuming flows \cite{citeulike:13587586},
and signals that measure the imbalance of the current liquidity. In \cite{leh16moun}, an example of how to use liquidity imbalance signals within a very short trading tactic was studied.
These two types of signals are closely related, since within short terms, price moves are driven by matching of liquidity supply and demand (i.e. current offers and consuming flows).

%I couldn't underand the following paragraph so I took it out%%%%%%%%%%%%%%%

%The dynamics of trading signals have mainly been indirectly studied by now. Academic studies are more focussed on liquidity dynamics (well documented in \cite{citeulike:6659908} and \cite{citeulike:8318790}) rather on providing models adequate to be used in optimal execution.
%A recognized feature linking flows and prices is the mean reverting nature of liquidity driven signals (see explanations in \cite{citeulike:12335801} or \cite{citeulike:12810809}).
%For this reason we use Ornstein-Uhlenbeck dynamics for the signal to provide illustrations in Section \ref{sec:ou}.

%%%%%%%%%%%%%%%%%%%%%%%%%%%%%%%%%%%%%%%%%%%%%%%%%%%%%%
As mentioned earlier, one of the major influencers on transaction costs is the market impact. Empirical studies have shown that the influence of the market impact is \emph{transient}, that is, it decays within a short time period after each trade (see \cite{citeulike:13497373} and references therein).
%on the short term price dynamics is the decay of the market impact. 
%We try to include this effect in our optimal trading framework too.
%It is why we choose to include a signal and market impact decay in the framework of Section \ref{sec-res}.
In this paper we will focus on two frameworks which take into account different types of market impact:
\begin{itemize}
\item Gatheral, Schied and Slynko (GSS) framework \cite{GSS}, in which the market impact is transient and strategies have a fuel constraint, i.e., orders are finished before a given date $T$;
% and its extension by Dang \cite{Dang2014}
\item Cartea and Jaimungal (CJ) framework \cite{CarteaJaimungal1}, where the market impact is instantaneous and the fuel constraint on the strategies is replaced by a smooth terminal penalization. 
\end{itemize}
%The GSS model has been partially extended in a working paper by Dang \cite{Dang2014}. Market impact for short time scales  was also studied by in \cite{citeulike:13990970}.
Note that \cite{GSS} is not the only framework with market impact decay. This kind of dynamics was originally introduced in \cite{Ob-Wan2005} and reused in \cite{AFS2} as in some other papers. We decided to focus on these two frameworks since they are extensively used in the financial literature. The model and analysis which are developed in this paper, could be applied also to other optimal trading frameworks.

The main theoretical result of this work deals with the addition of a Markovian signal into the optimal trading problem which was studied in \cite{GSS}. 
We will argue in Section \ref{sec:setup}, that this is modelled mathematically by adding a Markovian drift to the martingale price process. We formulate a cost functional which consists of the trading costs and the risk of holding inventory at each given time. 
Then we prove that there exists at most one optimal strategy that minimizes this cost functional. The optimal strategy is formulated as a solution to an integral equation. We then derive explicitly the optimal strategy, for the special case where the signal is an Ornstein-Uhlenbeck process. From the mathematical point of view this is the first time that a non martingale price process is incorporated to a optimal liquidation problem with a decaying market impact. Therefore the results of Theorems \ref{thm-uniq} and \ref{thm-exs}
extend Proposition 2.9 and Theorem 2.11 of \cite{GSS}, respectively.    
Later we show that in the asymptotic regime were the transient market impact becomes instantaneous, the singular optimal strategies which were derived in the (GSS) framework, becomes continuous. Moreover we show, that asymptotics of the optimal strategy in the (GSS) framework coincide with the optimal strategy which is obtained in the (CJ) framework (see Remark \ref{remark-limit} and Section 3).
This benchmark between different trading frameworks provides researchers and practitioners a wider overview, when they are facing realistic trading problems.
\medskip\\
The use of predictive signals optimal trading, in the context which was described above, is relatively new (see \cite{citeulike:13587586}). To the best of our knowledge, this is the first time that a Markovian signal and a transient market impact are confronted.
The (GSS) framework already includes a transient market impact, without using signals. The (CJ) framework includes only a bounded Markovian signal and not a decaying market impact. Moreover, our results on optimal trading in the (GSS) framework incorporate a risk eversion term to the cost functional, which was not taken into account in the results of \cite{GSS}.
\medskip \\
The main contribution of this work is in providing a new framework for optimal trading, which is  an extension of the classical frameworks of \cite{CarteaJaimungal1} and \cite{GSS} among others. The motivation to use this framework arises from market needs as our data analysis in Section \ref{sec:signal} suggests. From a theoretical point of view, these models of trading with signals provide some new mathematical challenges. We will describe in short two of these challenges. \medskip \\ 
The optimal strategies that we derive in Theorem \ref{thm-exs} and Corollary \ref{cor-cost} (i.e. in the GSS framework) are deterministic and they use only information on the signal at time $0$. One of the challenging questions which remain open, is how to optimise the trading costs over strategies which are adapted to the signal's filtration (see Remark \ref{rem-addp}).  \medskip  \\ 
An interesting phenomenon which arises from our results, is that the optimal strategies may not be monotone once we take into account trading signals (see figure \ref{fig:opttraj:fuelc}). It implies that price manipulations, triggered by trading strategies, are possible. Another challenge is to establish conditions on the market impact kernel function and on the signal, that will prevent price manipulations (see Remark \ref{rem-manip}).  

% Market impact decay has already been studied in the scope of this framework, like in others. This decay has been introduced in \cite{Ob-Wan2005} and reused in \cite{AFS2} as in some other papers.

Another contribution of this paper, is the statistical analysis of the imbalance signal and its use in actual trading, which we present in Section \ref{sec:signal}.
In order to validate our assumptions and theoretical results, we use nine months of real data from nordic European equity markets (the NASDAQ OMX exchange) to demonstrate the existence of a liquidity driven signal. We focus the analysis on 13 stocks, accounting for more than 9 billions of transactions. We also show that practitioners are at least partly conditioning their trading rate on this signal. Up to 2014, this exchange provided with each transaction the identity of the buyer and the seller. This database was already used for some academic studies, hence the reader can refer to \cite{citeulike:13497022} for more details. We added to these labelled trades, a database of Capital Fund Management (CFM) that contains information on the state of the order book just before each transaction. Thanks to this hybrid database, we were able to compute the imbalance of the liquidity just before decisions are taken by participants (i.e. sending a market orders which consume liquidity).

We divide most members of the NASDAQ OMX into four classes: global investment banks, institutional brokers, high frequency market makers and high frequency proprietary traders (the classification is detailed in the Appendix). Then, we compute the average value of the imbalance just before each type of participant takes a decision (see Figure \ref{fig:imb:otype}). The conclusion is that some participants condition their trading rate on the liquidity imbalance. Moreover, we provide a few graphs that demonstrate a positive correlation between the state of the imbalance and the future price move. These graphs also provide evidences for the mean-reverting nature of the imbalance signal (see Figure \ref{fig:preimb}--\ref{fig:preimb3}). In Figure \ref{fig:imbrate} we preset the estimated trading speed of market participants as a function of the average value of the imbalance, within a medium time scale of 10 minutes. The exhibited relation between the trading rate and the signal in this graph is compatible with our theoretical findings.

This paper is structured as follows. In Section \ref{sec-res} we introduce a model with a market impact decay, a Markovian signal and strategies with a fuel constraint (i.e in the (GSS) framework). We provide general existence and uniqueness theorems, and then give an explicit solution for the case of an Ornstein-Uhlenbeck signal. The addition of a signal to market impact decay is the central ingredient of this Section. In Section \ref{Sec-exmp} we compare our results from Section \ref{sec-res} to the corresponding results in the (CJ) framework. We  show that the optimal strategy in the (GSS) framework coincides with the optimal strategy in the (CJ) framework, in the asymptotic limit where the transient market impact become instantaneous and the signal is an Ornstein-Uhlenbeck process. 
In Section \ref{sec:signal} we provide an empirical evidence for the predictability of the imbalance signal and its use by different types of market participants. We also preform a statistical analysis which supports our focus on an Ornstein-Uhlenbeck signal in the example which is given in Section \ref{sec-res}. 
The last section is dedicated to proofs of the main results.

\section{Model Setup and Main Results} \label{sec-res}

\subsection{Model setup and definition of the cost functional}\label{sec:setup}
In this section we define a model which incorporates a Markovian signal into the (GSS) optimal trading framework. Definitions and results from \cite{GSS} are used throughout this section. 
 
We consider a probability space $(\Omega,\mathcal F, (\mathcal F_{t}),\mathbb P)$ satisfying the usual conditions, where $\mathcal F_{0}$ is trivial. Let $M=\{M_{t}\}_{t\geq 0}$ be a right-continuous martingale and $I=\{I_{t}\}_{t\geq 0}$ a homogeneous c\`adl\`ag Markov process satisfying,
\be
\int_{0}^{T}E_{\iota}\big[ |I_{t}|\big]\,dt <\infty, \quad \textrm{for all } \iota \in \re, \ T>0. 
\ee
Here $E_{\iota}$ represents expectation conditioned on $I_{0}=\iota$.  
In our model $I$ represents a signal that is observed by the trader. 

We assume that the asset price process $P$, which is unaffected by trading transactions, is given by
\bn
dP_{t}= I_{t}dt + dM_{t}, \quad t\geq 0,
\en  
hence the signal interacts with the price through the drift term. This setting allows us to consider a large class of signals. 
The visible asset price, which is described later, also depends on the market impact that is created by trader's transactions. 

Let $[0,T]$ be a finite time horizon and $x>0$ be the initial inventory of the trader. Let $X_{t}$ be the amount of inventory held by the trader at time $t$. We say that $X$ is an admissible strategy if it satisfies:
\begin{itemize} 
\item [\textbf{(i)}]  $t\longrightarrow X_{t}$ is left--continuous and adapted. 
\item [\textbf{(ii)}] $t\longrightarrow X_{t}$ has finite and $\mathbb{P}$-a.s. bounded total variation. 
\item [\textbf{(iii)}] $X_{0}=x$ and $X_{t}=0$, $\mathbb{P}$-a.s. for all $t>T$. 
\end{itemize} 
As in \cite{GSS,Dang2014,Cur-Gath17}, we assume that the visible price $S=\{S_{t}\}_{t\geq0 }$ is affected by a transient market impact, and it is given by 
\be \label{market-imp}
S_{t} = P_{t}+\int_{\{s<t\}}G(t-s)dX_{s}, \quad t\geq 0, 
\ee
where the \emph{decay kernel} $G:(0,\infty) \rightarrow  [0,\infty)$ is a measurable function such that the following limit exists
\be \label{g-0}
G(0):=\lim_{t\downarrow 0}G(t).
\ee  
 
Next we derive the transaction costs which are associated with the execution of a strategy $X_{t}$.  

Note that if $X_{t}$ is continuous in $t$, then the trading costs that arise by an infinitesimal order $dX_{t}$ are $S_{t}dX_{t}$. When  $X_{t}$ has a jump of size $\Delta X_{t}$ at $t$, the price moves from $S_{t}$ to $S_{t+}=S_{t}+G(0)\Delta X_{t}$ and the resulted costs by the trade $\Delta X_{t}$ are given by (see Section 2 of \cite{GSS}) 
\bn
\frac{G(0)}{2}\big(\Delta X_{t}\big)^{2} + S_{t}\Delta X_{t}.
\en
It follows that the trading costs which arise from the strategy $X$ are given by 
\bn
\int S_{t}dX_{t} +\frac{G(0)}{2}\sum\big(\Delta X_{t}\big)^{2} &=&\int \int_{0}^{t}I_{s}\,ds\,dX_{t}+\int\int_{\{s<t\}}G(t-s)dX_{s}dX_{t} \\
&&\quad +\int M_{t}dX_{t} +\frac{G(0)}{2}\sum\big(\Delta X_{t}\big)^{2}.
\en
From Lemma 2.3 in \cite{GSS}, we get a more convenient expression for the expected trading costs,   
\bn
&&E\Big[\int \int_{0}^{t}I_{s}\,ds\,dX_{t}+\int \int_{\{s<t\}}G(t-s)dX_{s}dX_{t}  +\int M_{t}dX_{t} +\frac{G(0)}{2}\sum\big(\Delta X_{t}\big)^{2}\Big] \\
&&=E\Big[\int \int_{0}^{t}I_{s}\,ds\,dX_{t}+\frac{1}{2} \int \int G(|t-s|)dX_{s}dX_{t}\Big] -P_{0}x.
\en 
We are interested in adding a risk aversion term to our cost functional. A natural candidate is $\int_0^TX_t^2\,dt$, which is considered as a measure for the risk associated with holding a position $X_t$ at time $t$; see~\cite{AlmgrenSIFIN,Forsythetal,Tseetal}
and the discussion in Section 1.2 of~\cite{SchiedFuel}. 
Hence our cost functional which is the sum of the expected trading costs and the risk aversion term has the form 
\be \label{ex-cost}
 E\Big[\int \int_{0}^{t}I_{s}\,ds\,dX_{t}
+\frac{1}{2}\int \int G(|t-s|)dX_{s}dX_{t}+\phi\int_0^TX_t^2\,dt\Big]-P_{0}x, 
\ee
where $\phi\geq 0$ is a constant. 
 
The main goal of this work is to minimize this cost functional (\ref{ex-cost}) over the class of admissible strategies. 
Before we discuss our main results in this framework, we introduce the following class of kernels. 

We say that a continuous and bounded $G$ is strictly positive definite if for every measurable strategy $X$ we have 
\be \label{pos-def}
\int \int G(|t-s|)\,dX_s\,dX_t >0, \quad P-\rm{a.s.}
\ee
We define $\mathbb G$ to be the class of continuous, bounded and strictly positive definite functions $G:(0,\infty)\rr [0,\infty)$. 
\begin{remark} 
Note that (\ref{g-0}) is satisfied for every $G\in \mathbb G$. A characterization of positive definite kernels (that is, when the inequality (\ref{pos-def}) is not strict) is given in Proposition 2.6 in \cite{GSS}. 
\end{remark}
\begin{remark} 
An important subclass of $\mathbb{G}$ is the class of bounded, non increasing convex functions $G:(0,\infty) \rr [0,\infty)$ (see Proposition 2 in \cite{alf-sch12}). 
\end{remark}

\subsection{Results for a Markovian Signal}\label{sec:results}
In this section we introduce our results on the existence and uniqueness of an optimal strategy, when the signal is a  c\`adl\`ag Markov process. 
As in Section 2 of \cite{GSS} we restrict our discussion to deterministic strategies.  The minimization of the cost functional over signal-adaptive random strategies, will be discussed in Remark \ref{rem-addp}.

We consider the following class of strategies, 
\begin{equation*} 
\Xi(x) = \{ X | \textrm{ deterministic admissible strategy with } X_{0}=x \ \textrm{and support in  }  [0,T]  \}. 
\end{equation*}
Note that for any $X$ in $\Xi(x)$, the cost functional (\ref{ex-cost}) has the form, 
\be \label{ex-cost2}
\int \int_{0}^{t}E_{\iota}[I_{s}]\,ds\,dX_{t}
+\frac{1}{2}\int \int G(|t-s|)dX_{s}dX_{t}+\phi\int_0^TX_t^2\,dt.
\ee
In our first main result we prove that there exists at most one strategy which minimizes the cost functional (\ref{ex-cost2}). 
\begin{theorem} \label{thm-uniq}
Assume that $G \in \mathbb G$. Then, there exists at most one minimizer to the cost functional (\ref{ex-cost2}) in the class of admissible strategies $\Xi(x)$.  
\end{theorem} 
In our next result we give a necessary and sufficient condition for the minimizer of the cost functional (\ref{ex-cost2}). 
\begin{theorem}  \label{thm-exs}
$X^{*} \in \Xi(x)$ minimizes the cost functional (\ref{ex-cost2}) over $\Xi(x)$, if and only if there exists a constant $\lambda$ such that $X^{*}$ solves 
\begin{equation}\label{cond-opt} 
\int_{0}^{t}E_{\iota}[I_{s}]ds+ \int G(|t_{}-s|)dX^{*}_{s}-2\phi\int_{0}^{t}X^{*}_{s}ds =\lambda, \quad \textrm{for all } 0\leq t\leq T. 
\end{equation}
\end{theorem} 
A few remarks are in order. 
\begin{remark} 
In the special case where the agent does not rely on a signal (i.e. $I=0$) and there is a zero risk aversion ($\phi =0$), Theorems 
(\ref{thm-uniq}) and (\ref{thm-exs}) coincide with Proposition 2.9 and Theorem 2.11 in \cite{GSS}. 
\end{remark}
\begin{remark} 
Dang studied the case where the risk aversion term in (\ref{ex-cost}) is nonzero, but again $I=0$. In Section 4.2 of \cite{Dang2014}, a necessary condition for the existence of an optimal strategy is given, when the admissible strategies are deterministic and absolutely continuous. Our condition in (\ref{cond-opt}) coincides with Dang's result when $I=0$ and the admissible strategies are absolutely continuous. Note however that the question if the condition in \cite{Dang2014} is also sufficient and the uniqueness of the optimal strategy, remained open even in the special case where $I=0$. 
\end{remark}

\subsection{Result for an Ornstein-Uhlenbeck Signal}\label{sec:ou}
As mentioned in the introduction, a special attention is given to the case where the signal $I$ is an Ornstein--Uhlenbeck
process,    
\be
\begin{aligned} \label{I-OU} 
dI_{t}&= -\gamma I_{t}\, dt +\sigma \, dW_{t},    \quad t\geq 0, \\
I_{0}&=\iota, 
\end{aligned} 
\ee
where $W$ is a standard Brownian motion and  $\gamma, \sigma>0$ are constants. In the following corollary we derive an explicit formula for the optimal strategy in the case of zero risk aversion and when $G$ has an exponential decay. The following corollary generalizes the result of Obizhaeva and Wang \cite{Ob-Wan2005}, who solved this control problem when there is no signal.
\begin{corollary}  \label{cor-cost}
Let $I$ be defined as in (\ref{I-OU}). Assume that $\phi =0$ and $G(t) = \kappa\rho e^{-\rho t}$, where $\kappa, \rho >0$ are constants. Then, there exists a unique minimizer $X^{*}\in\Xi(x)$ to the cost functional (\ref{ex-cost2}), which is given by 
 \be \label{opt-spec}
X^{*}_{t}=(1-b_0(t))\cdot x + {\iota \over 2\kappa\rho^2\gamma}\left\{ \frac{\rho^2-\gamma^2}{\gamma} \cdot b_1(t) - (\rho + \gamma) \cdot b_2(t) - (\rho + \gamma) \cdot b_3(t)\right\},
% X^{*}_{t}=x+\one_{\{t>0\}}A+\frac{B}{\gamma}\big(1-e^{-\gamma t}\big)+Ct+ \one_{\{t>T\}}D.
 \ee
where
  \bn
b_0(t) &=& {\one_{\{t>0\}}+\one_{\{t>T\}}+ \rho t\over 2 + \rho T},\\
b_1(t) &=&1 - e^{-\gamma t} -b_0(t)(1-e^{-\gamma T}),\\
b_2(t) &=&\one_{\{t>T\}}+ \rho t - b_0(t)(1+\rho T),\\ 
b_3(t) &=&(b_0(t) - \one_{\{t>T\}}) e^{-\gamma T}.
%  A&=&\frac{1}{2+T\rho}\Big(\frac{\iota}{2\kappa \rho^{2}\gamma}\Big((\rho+\gamma)\big(1+T\rho-\gamma^{-1}(\rho-\gamma)(1-e^{{-\gamma T}})\big)-(\rho-\gamma)e^{-\gamma T}\Big)-x\Big), \\
%  B&=&\iota \frac{\rho^2-\gamma^{2}}{2\kappa \rho^{2} \gamma}, \\
% C&=&\rho A-\iota\frac{\rho+\gamma}{2\kappa \rho \gamma}, \\
% D&=&A-\frac{\iota}{2\kappa \rho^{2}\gamma}((\rho+\gamma)-(\rho-\gamma)e^{-\gamma T}). 
 \en 
% Note that $A,C,D$ are functions of $(x,\iota,T)$ while $B$ is a function of $\iota$.
 \end{corollary} 
Note that $1-b_0(T+)=0$ and $b_{i}(T+)=0$ for $i=1,2,3$, moreover, the optimal strategy is linear both in $x$ and $\iota$.

 \begin{figure}[ht!] 
   \centering
   \includegraphics[width=.9\linewidth]{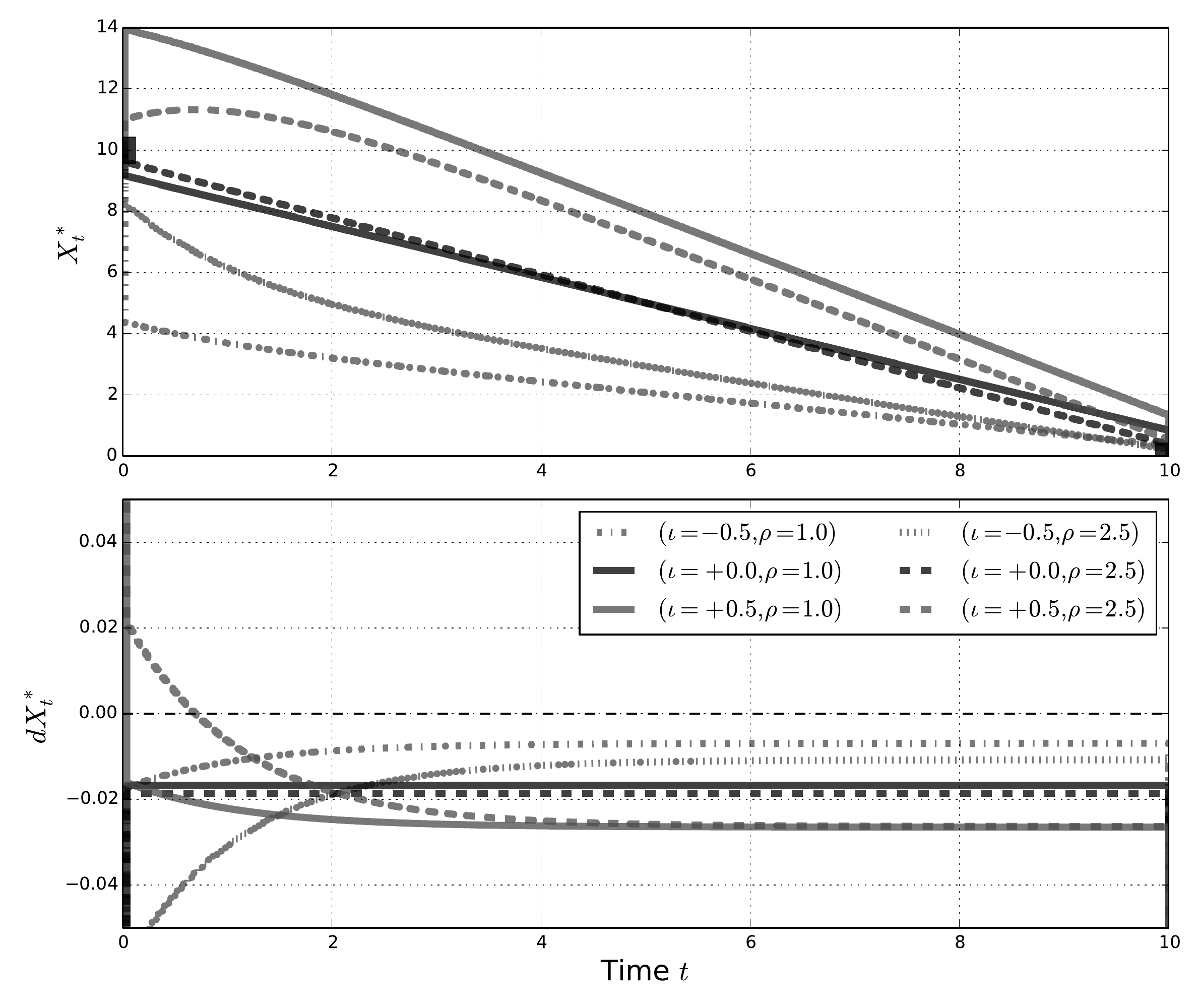}
   \caption{Optimal trading strategies according to (\ref{opt-spec}) for $\gamma = 0.9, \ \kappa = 0.1, \  T =  10$ and $x  = 10$.  We demonstrate different scenarios for selling $10$ shares: without a signal, with a positive signal and with a negative signal. We distinct between a slow decay of the market impact (solid lines) or fast decay (dashed lines).
At the top graph we show the remaining inventory, at the bottom graph the trading speed (for $0<t<10$) is presented.}
   \label{fig:opttraj:fuelc}
 \end{figure}

In Figure \ref{fig:opttraj:fuelc} we present some examples of optimal strategy with the following parameters: 
$\gamma = 0.9, \kappa = 0.1, T =  10, x    = 10$.
These particular values are compatible with the empirical parameters which are estimated at the end of Section \ref{sec:imb:signal}. Arbitrary initial values (-0.5, 0 and +0.5) are taken for the signal $\iota$. The special case where $\iota=0$ gives similar result to Obizhaeva and Wang \cite{Ob-Wan2005}. The parameter $\rho$, which controls the market impact decay, cannot be estimated from the data that we have, hence we take two arbitrary but realistic values (1.0 and 2.5). We observe that for large values of $\rho$, the initial jump in the optimal trading strategy is larger than the corresponding jump in the small $\rho$ strategies, but the trading speed tends to have a less variations. 
We particularly notice that when the initial signal is at an opposite direction to the trading ($\iota >0$ for a sell order) the trading starts with purchases as expected, and afterwords the trading speed eventually becomes negative.  On the other hand, when the initial signal is at the same direction to the trading, it is optimal to start selling immediately and most of the inventory is sold before $T/2$.

%\begin{table}[ht!]
%  \centering
%\begin{tabular}{rrr}
%\toprule
%{} &     $\rho$ &        \\
%$\iota$   &    1.0 &    2.5 \\
%\midrule
%-0.5 &  2.118 &  3.582 \\
% 0.0 & -0.014 & -0.008 \\
% 0.5 & -7.052 & -6.058 \\
%\bottomrule
%\end{tabular}  \caption{Value of the optimal strategie (the has to be minimized) for different initial values of the signal $\iota$ and for different decays $\rho$.}
 % \label{tab:values:i:r}
%\end{table}

%Table \ref{tab:values:i:r} shows the asymmetry in term of value function (keep in mind the value function has to be minimized) of having a good or bad signal for the optimal strategy: the gain is larger (7.1 or 6.1) when the signal is positive, than the lost (2.1 or 3.5) when the signal is negative (with the same amplitude). It is due to the fuel constraint to have zero inventory at $T$.

 In the following remarks we discuss the result of Corollary \ref{cor-cost}. 
 
\begin{remark}  \label{remark-limit} 
Note that in the limit where $\rho \rr \infty$, the market impact term in (\ref{ex-cost2}), $\frac{1}{2}\int_{0}^{T}\int_{0}^{T}G(|t-s|)dX_{s}dX_{t}$ formally corresponds to the costs arising from instantaneous market impact, that is $G(dt)= \kappa\delta_{0}$. We briefly discuss  the asymptotics of the optimal strategy $X^{*}_{t}= X^{*}_{t}(\rho)$ in (\ref{opt-spec}) when $\rho \rr \infty$. It is easy to verify that in the limit, the jumps of $X^{*}$ vanish (see $A$ and $D$ in (\ref{constants}) for the explicit expression of the jumps), and the limiting optimal strategy $X^{*}(\infty)$ is a smooth function which is given by 
\bn
X^{*}_{t}(\infty) = X+\frac{\iota}{2\kappa \gamma^{2}}\big(1-e^{-\gamma t}\big)-\frac{\iota}{2\kappa \gamma }t. 
\en
Motivated by these asymptotic results, in the next section we further explore absolutely continuous strategies which minimize the trading costs -- risk aversion functional. We will assume there that the market impact is instantaneous, that is $G(dt)= \kappa\delta_{0}$ and drop the fuel constraint ($X_{t}=0$ for $t>T$) from the admissible strategies. Then, explicit formulas for the optimal strategy are derived when the risk aversion term is non-zero. 
\end{remark} 

\begin{remark}[An adaptive version of (\ref{opt-spec})] \label{rem-addp}
  Equation (\ref{opt-spec}) gives an optimal solution for a trader with inventory $X_0=x$ at $t=0$, who is observing the initial value of the signal $\iota=I_0$ and wishes to minimize (\ref{ex-cost2}) for exponentially decaying kernel  and $\phi =0$. The cost functional is there given by, 
  \begin{equation*}
    \label{eq:spec-cost}
    U([0,T]) :=   \int_{\tau\in [0,T]} \int_{s\in [0,\tau]}E[I_{s}|  \mathcal F^{W}_0] \,ds\,dX_{\tau} +\frac{1}{2} \int_{\tau\in [0,T]} \int_{s\in [0,T]} \kappa \rho e^{-\rho |\tau-s|}dX_{s}dX_{\tau}, 
  \end{equation*}
where $(\mathcal F^{W}_t)_{t\geq 0}$ is the natural filtration of $I$. 
In this setting, once the trading started, it is no longer possible to update the strategy while taking into account new information, i.e. new values of the signal. 
 This can be compared to simpler frameworks like the one of Section \ref{Sec-exmp}, in which the optimal strategy is updated for any $0\leq t\leq T$. 
   \def\tX{{\tilde X}}\def\tU{{\tilde U}}
We therefore add a short discussion on an adaptive framework for (\ref{opt-spec}).

A natural way to update the optimal strategy at any time $t$, is to define $\tX:=\{\tX_{s}: t\leq s \leq T\}$ as the optimal strategy of the cost functional  
  $$\wt U([t,T]) :=  \int_{\tau \in [t,T]} \int_{s\in [t,\tau]}E[I_{s}|\mathcal F^{W}_t] \,ds\,dX_{\tau} +\frac{1}{2} \int_{\tau\in[t,T]} \int_{s\in [t,T]} \kappa \rho e^{-\rho |\tau-s|}dX_{s}dX_{\tau} .$$
Note however that 
  $$U([0,T]) = U([0,t]) + \Delta_1 U(t,T)+\Delta_2 U(t,T),$$ 
  where 
$$\Delta_1 U(t,T) = \int_{\tau\in[t,T]} \int_{s\in [0,\tau]}E[I_{s}|\mathcal F_{0}^{W}]\,ds\,dX_{\tau}  +\frac{1}{2} \int_{\tau \in [t,T]} \int_{s\in[0,t]}\kappa \rho e^{-\rho |\tau-s|}dX_{s}dX_{\tau}$$
and 
$$\Delta_2 U(t,T) =\frac{1}{2} \int_{\tau \in [0,T]} \int_{s\in [t,T]} \kappa \rho e^{-\rho |\tau-s|}dX_{s}dX_{\tau}.$$ 
This implies \emph{if $\tX$ is used in place of $X^*$ for some $\tau \in(t,T)$, the trader will have an $\mathcal F_{t}$-adapted control , but it will not be necessarily consistent with $X^{*}$ which minimizes $U([0,T])$}. Therefore, in practice one can choose between the following options:
\begin{itemize}
\item the optimal strategy $X^*$, limited to the information on the signal at $t=0$;
\item an approximate strategy $\tX$ updated at each time $t\in(0,T)$, which takes into account the whole trajectory of $I_t$;
\item the optimal strategy which corresponds to a market impact without a decay (as shown in Section \ref{Sec-exmp}). 
\end{itemize}
The question which of these strategies gives the best results remains open. 
  
Note that in the cost functional $U$ the time inconstancy is a result of the transient  market impact term. In \cite{schon}, time inconsistent optimal liquidation problems were  also studied. However, the inconstancy of the problems in \cite{schon} arises from the risk-aversion term.
\end{remark} 
\begin{remark}[Price manipulation]  \label{rem-manip}
Market impact models admit transaction-triggered price manipulation if the expected costs of a sell (buy) strategy can be reduced by intermediate buy (sell) trades (see Definition 1 in \cite{alf-sch12}).  Theorem 2.20 in \cite{GSS} implies that transaction-triggered price manipulation are impossible for the cost functional in \ref{cond-opt}, over the class of admissible strategies, in the case where $I\equiv 0$ and $\phi =0$.  However, Figure \ref{fig:opttraj:fuelc}, shows that adding signals to the same market impact model can create optimal strategies which are not monotone decreasing, and therefore implies a possible price manipulation. 
It would be very interesting to investigate the conditions on the market impact kernel and the trading signals which ensure no price manipulations. Study of the possible implications of these price manipulations on other market participants is also of major importance.
 \end{remark}

\section{Optimal strategy for temporary market impact}  \label{Sec-exmp}
In this section we study an optimal trading problem which has some common features to the problem which was introduced in Section \ref{sec:setup}. We consider again a price process which incorporates a Markovian signal. The main change in this section, is that the market impact in (\ref{market-imp}) is temporary, i.e. the kernel is given by $G(dt)= \kappa\delta_{0}(dt)$, where $\delta_{0}$ is Dirac's delta measure and $\kappa>0$ is a constant. Note that this type of kernel is not included in the class of kernels $\mathbb G$ which was introduced in (\ref{market-imp}). The main goal of this section is to show of how to incorporate trading signals in the (CJ) framework \cite{CarteaJaimungal1}. The results that we obtain could be compared to the results of Section \ref{sec-res} (see Remark \ref{remark-limit}). Recall that we heuristically obtained the optimal strategy when the kernel $G=G_{\rho}$ ``converges'' to Dirac's delta measure as $\rho \rightarrow 0$.

 We continue to assume that $I$ is a  c\`adl\`ag Markov process as in the beginning of Section \ref{sec-res} but we add the assumption that 
\be \label{I-asmp} 
E_{\iota}\big[|I_{t}|] \leq C(T)(1+|\iota|), \quad \textrm{for all } \iota \in \re,  \ 0\leq t\leq T, 
\ee 
for some constant $C(T)>0$. 
%Note that (\ref{I-asmp}) can be easily verified for a large class of diffusions. 
%A typical example is when the drift and noise coefficients of the diffusion has at most linear growth.

For the sake of simplicity we will assume that $M_{t}=\sigma^{P}W_{t}$ so that 
\bn
dP_{t}=I_{t}dt+\sigma^{P}dW_{t}, 
\en
where $\{W_{t}\}_{t\geq 0}$ is a Brownian motion and $\sigma^{p}$ is a positive constant. 

In the following example the fuel constraint on the admissible strategies will be replaced with a terminal penalty function. This allows us to consider absolutely continuous strategies as in the framework of Cartea and Jaimungal  (see e.g. \cite{Car-Jiam-2013, Car-Jiam15, Car-Jiam-2016}). We introduce some additional definitions and notation which are relevant to this setting.  

Let $\mathcal V$ denote the class of progressively measurable control processes $r=\{r_{t}\}_{t\geq0}$ for which 
$\int_{0}^{T}|r_{t}|dt<\infty$, $P$-a.s. 

For any $x \geq 0$ we define 
\be \label{inv-eq}
X_{t}^{r}=x-\int_{0}^{t}r_{t}dt. 
\ee
Here $X^{r}_{t}$ is the amount of inventory held by the trader at time $t$. We will often suppress the dependence of $X$ in $r$, to ease the notation. 

The price process, which is affected by the linear instantaneous market impact, is given by
\bn
S_{t} = P_{t}-\kappa r_{t}, \quad t\geq 0, 
\en
where $\kappa>0$. Note that $S_{t}$ here corresponds to (\ref{market-imp}) when $G(dt)=\kappa\dl_{0}(dt)$.

The investor's cash $\mathcal C_{t}$ satisfies
\bn
d\mathcal C_{t} := S_{t}r_{t}\,dt =(P_{t}-\kappa r_{t})r_{t}\,dt, 
\en
with $\mathcal C_{0} =c$. 

For the sake of consistency with earlier work of Cartea and Jaimungal in \cite{Car-Jiam-2013, Car-Jiam15, Car-Jiam-2016}, we will define the liquidation problem as a maximization of the difference between the cash and the risk aversion. As mentioned earlier,  the fuel constraint on the admissible strategies will be replaced by the penalty function, which is given by $X_{T}(P_{T}-\varrho X_{T})$ where $\varrho$ is a positive constant. 

The resulted cost functional is given by   
\begin{equation}  \label{v-cost} 
V^{r}(t,\iota,c,x,p) =E_{\iota,c,x,p}\Big[\mathcal C_{T}-\phi\int_{t}^{T}X^{2}_{s}ds+X_{T}(P_{T}-\varrho X_{T})\Big],
\end{equation} 
where $\phi \geq 0$ is a constant and $E_{t,\iota,c,x,p}$ represents expectation conditioned on $I_{t}=\iota, \mathcal C_{t} =c, X_{t}=x, P_{t}=p,$.  

The value function is   
\bn
V(t,\iota,c,x,p)=\sup_{r\in\mathcal V}V^{r}(t,\iota,c,x,p).
\en
Note that this control problem could be easily transformed to a minimization of the trading costs and risk aversion as in Section \ref{sec-res}.

Let $\mathcal L^{I}$ be the generator of the process $I$. Then, the corresponding HJB equation is 
\begin{align}  \label{hjb} 
0&=\partial_{t} V+ \iota \partial_{p} V+\frac{1}{2}(\sigma^P)^{2}\partial^{2}_{p}V+\mathcal L^{I}V-\phi x^{2}+\sup_{r }\Big\{r\big(p-\kappa r\big)\partial_{c}V-r \partial_{x}V\Big\},
\end{align} 
with the terminal condition 
$$
V(T,\iota,c,x,p)= c+x(p-\varrho x).
$$
Let $E_{t,\iota}$ represent expectation conditioned on $I_{t}=\iota$. 
In the following proposition we derive a solution to (\ref{hjb}). The proof of Proposition \ref{prop-hjb} follows the same lines as the proof of Proposition 1 in \cite{Car-Jiam-2016}. 
\begin{proposition} \label{prop-hjb} 
Assume that $\varrho \not = \sqrt{\kappa \phi}$. Then, there exists a solution to (\ref{hjb}) which is given by 
\be \label{opt-V}
V(t,\iota,c,x,p)=c-xp+ v_{0}(t,\iota) + xv_{1}(t,\iota)+x^{2}v_{2}(t), 
 \ee
 where 
 \bn
 v_{2}(t) &=& \sqrt{\kappa \phi} \frac{1+\zeta e^{2\beta(T-t)}}{1-\zeta e^{2\beta(T-t)}}, \\
 v_{1}(t,\iota)&=&\int_{t}^{T}e^{\frac{1}{\kappa}\int_{t}^{s}v_{2}(u)du} E_{t,\iota}[I_{s}]ds, \\
 v_{0}(t,\iota)  &=&\frac{1}{4\kappa}\int_{t}^{T}E_{t,\iota}\big[v^{2}_{1}(s,I_{s})\big]ds, 
 \en
and the constants  $\zeta$ and $\beta$ are given by 
\bn
\zeta = \frac{\varrho+ \sqrt{\kappa \phi}}{\varrho- \sqrt{\kappa \phi}}, \qquad  \beta = \sqrt{\frac{\kappa}{ \phi}}.
\en
\end{proposition}   
In the following proposition we prove that the solution to (\ref{hjb}) is indeed a an optimal control to (\ref{v-cost}). 
\begin{proposition} \label{verification} 
Assume that $\varrho \not = \sqrt{\kappa \phi}$. Then (\ref{opt-V}) maximizes the cost functional in (\ref{v-cost}). The optimal trading speed $r_{t}^*$ is given by 
\bn
r^{*}_{t}=  -\frac{1}{2\kappa}\Big(2v_{2}(t)X_{t}+\int_{t}^{T}e^{\frac{1}{\kappa}\int_{t}^{s}v_{2}(u)du} E_{t,\iota}[I_{s}]ds\Big), \quad 0\leq t\leq T.
\en
\end{proposition}  
The proofs of Propositions \ref{prop-hjb} and \ref{verification} are given in Section \ref{Sec-hjb}.  \medskip \\
The following corollary follows directly from Propositions \ref{prop-hjb} and \ref{verification}. Note that an Ornstein-Uhlenbeck process satisfies (\ref{I-asmp}).
\begin{corollary} \label{corr-smth}
Assume the same hypothesis as in Proposition \ref{prop-hjb}, only now let $I$ follow an Ornstein-Uhlenbeck process as in (\ref{I-OU}). Then, there exists a maximizer $r^{*}\in \mathcal V$ to $V^{r}(t,\iota,c,x,p)$, which is given by 
\bn
r^{*}_{t}=  -\frac{1}{2\kappa}\Big(2v_{2}(t)X_{t}+I_{t}\int_{t}^{T}e^{-\gamma (s-t)+ \frac{1}{\kappa}\int_{t}^{s}v_{2}(u)du}ds\Big), \quad 0\leq t\leq T.
\en
\end{corollary}
In the following remarks we compare between the results of Sections \ref{sec-res} and \ref{Sec-exmp}. 
\begin{remark} 
If we set the risk aversion and penalty coefficients $\phi, \varrho$ in (\ref{v-cost}) to $0$, then from the proof of Proposition \ref{prop-hjb} it follows that $v_{2}\equiv 0$. Under the same assumptions on the signal as in Corollary \ref{corr-smth}, the optimal strategy is given by
\be \label{uo-cj}
r^{*}_{t}=  -\frac{I_{t}}{2\kappa \gamma}\big(1-e^{-\gamma(T-t)}\big)  , \quad 0\leq t\leq T, 
\ee 
which is consistent with $X_{t}^{*}(\infty)$ from Remark \ref{remark-limit}.   
\end{remark} 
\begin{remark} 
One can heuristically impose a ``fuel constraint'' on the optimal strategy in Corollary \ref{corr-smth}, by using the asymptotics of $r_{t}^{*}$ when $\varrho \rr \infty$. In this case $\zeta \rr 1$ and the limiting optimal speed which we denote by $r_{t}^{f}$ is 
\bn
r^{f}_{t}=  -\frac{1}{2\kappa}\Big(2\bar v_{2}(t)X_{t}+I_{t}\int_{t}^{T}e^{-\gamma (s-t)+ \frac{1}{\kappa}\int_{t}^{s}\bar v_{2}(u)du}ds\Big), \quad 0\leq t\leq T.
\en
where
\bn
\bar v_{2}(t) &=&- \sqrt{\kappa \phi} \frac{1+ e^{2\beta(T-t)}}{1- e^{2\beta(T-t)}}.
\en
\end{remark} 
\begin{remark} 
It is important to notice that (\ref{opt-spec}) gives the optimal strategy on the time horizon $[0,T]$ in the (GSS) framework, by using only information on the O.U. signal at $t=0$. On the other hand (\ref{uo-cj}), which is the optimal trading speed $r_{t}$ in the (CJ) framework is using the information on the signal at time $t$. A crucial point here is, that if one tries to solve repeatedly the control problem in the (GSS) framework on time intervals $[t,T]$ for any $t>0$, by using $I_{t}$ and $S_{t}$ as an input, the optimal strategy will not necessarily minimize the cost functional (\ref{ex-cost}) on $[0,T]$. The reason for that is that the control problem in (\ref{ex-cost}) may be is inconsistent. The market impact (and therefore the transaction costs) which are created on $[0,t]$ affect the cost functional at $[t,T]$ (see Remark \ref{rem-addp} for more details). This effect disappears when we set $G(dt)=\dl_{t}$ in (\ref{v-cost}). 
\end{remark} 
In Figure 2 we simulate the optimal inventory $X^{*}$ which is resulted by the optimal  trading speed $r^{*}$ from Corollary \ref{corr-smth}. In the black solid line we present the optimal inventory in the case where there is no signal, therefore in this case the optimal strategy is deterministic. The red region in Figure 2 is a ``heat map'' of $1000$ realizations of the optimal inventory $X^{*}$. The pamperers of the model are $\gamma=0.1$, $\sigma =0.1$ and $I_{0}=0$ for I as in (\ref{I-OU}), and $T=10$, $\kappa=0.5$, $\phi=0.1$, $X_{0}=10$ and $\varrho = 10$, for the cost functional (\ref{v-cost}).  We observe that the random strategies are a perturbation of the classical deterministic optimal strategy. 
In Figure \ref{value-pic} we present the value function (\ref{opt-V}) at $t=0$, under the same assumptions as in Figure 2, that is, assuming that $I$ is an OU process and that the model parameters are similar. More precisely, we plot $V(0,\iota,c,x,p)-(c-xp)$, hence we omit constants which do not contribute to the behaviour of the model. We observe that the revenue resulted by the optimal sell strategy $r^{*}$ is affected by the direction and value of the signal $\iota$. The revenue of a sell strategy when the signal positive, which indicates on a potential price increase, is higher than negative signal scenarios.

\begin{figure}
\centering 
\includegraphics[width=10cm]{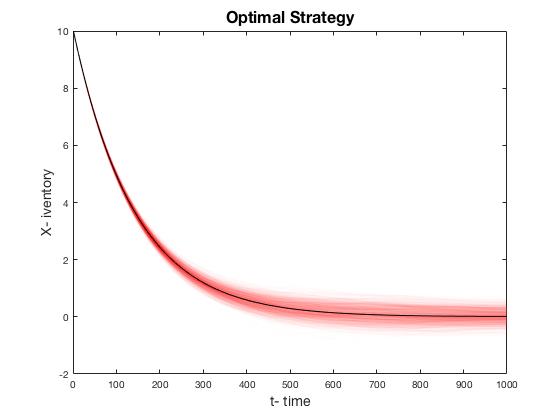} 
\caption{Simulation of the optimal inventory $X^{*}$ which is resulted by the trading speed $r^{*}$ from Corollary \ref{corr-smth}. In the black curve we present the optimal inventory in absence of a signal. The red region is a plot of $1000$ trajectories of the optimal inventory $X^{*}$. The pamperers of the model: $\gamma=0.1$, $\sigma =0.1$, $I_{0}=0$, $T=10$, $\kappa=0.5$, $\phi=0.1$, $X_{0}=0$ and $\varrho = 10$.} 
\end{figure}   \label{inv-pic}

\begin{figure} 
\centering
\includegraphics[width=10cm]{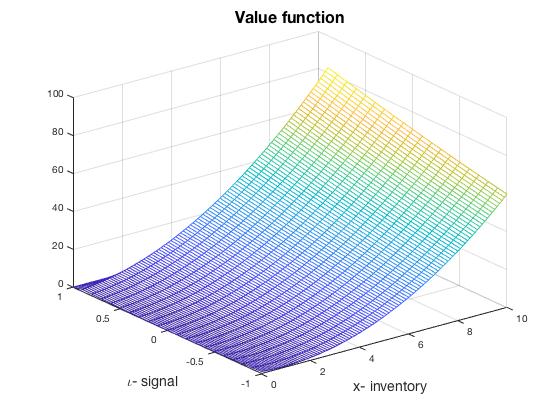}
\caption{Plot of the value function $V(0,\iota,c,x,p)-(c-xp)$ from (\ref{opt-V}), when the signal is an OU process. The pamperers of the model: $\gamma=0.1$, $\sigma =0.1$, $T=10$, $\kappa=0.5$, $\phi=0.1$, $X_{0}=0$ and $\varrho = 10$.}
 \end{figure}\label{value-pic}

\section{Evidence for the use of signals in trading} \label{sec:signal}
In this section we analyse financial data which is related to the limit order book imbalance. The data analysis in this section is directed to support the models which were introduced in Sections \ref{sec-res} and \ref{Sec-exmp}. 
In Section \ref{sec-data-base} we describe our data base and provide an empirical evidence for the use of the imbalance signal, which is a liquidity driven signal. In Section \ref{sec:imb:signal} we study the statistical properties of the signal and motivate  our model from Section \ref{sec:ou}, of an Ornstein-Uhlenbeck Signal. Finally in Section \ref{sec-use-sig} we study the use of this signal during liquidation by different market participants. 
Note that in Sections \ref{sec-res} and \ref{Sec-exmp} we also discussed more general signals which are not necessarily liquidity driven.

\subsection{The database: NASDAQ OMX trades}  \label{sec-data-base}

The database which is used in this section is made of transactions on NASDAQ OMX exchange. This exchange used to publish the identity of the buyer and seller of each transaction until 2014.
To obtain order book data, we use recordings made by Capital Fund Management (CFM) on the same exchange, which were matched with NASDAQ OMX trades thanks to the timestamp, quantity and price of each trade. On a typical month, the accuracy of such matching is more than 99.95\%.

The NASDAQ OMX trades were already used for academic studies (see
\cite{citeulike:13497022} and \cite{leh16moun} for details). We study 13 stocks traded on NASDAQ OMX Stockholm from January 2013 to September 2013.
The purpose of this section is not to conduct an extensive econometric study on this database; such work deserves a paper of its own. Our goal here is to show qualitative evidences for the existence of the order book imbalance signal and to study how market participants decisions depend on its value.

The 13 stocks which are used in this section have been selected for this research, since High Frequency Proprietary Traders took part in at least 100,000 trades on each of them during the studied period. More details on the classification of the traders to different classes are given later in this section.

\def\redsize{\small}
\begin{table}[ht!]
  \centering
\hspace*{-2em}{\small
\begin{tabular}{lrrrrrr}
\toprule
Company & Daily & Avg.  &   Avg.    & Volatility & Min. \\
Name (code) & Traded    & Price & BA-spread & (GK)       & tick \\
            & Value ($10^6$) &        &                & \\
\midrule
\redsize Volvo AB ({\tiny VOLVb.ST})                        &          431.20 &     94.87 &           0.057 & 15.08\% &    0.05 \\
\redsize Nordea Bank AB ({\tiny NDA.ST})                    &          384.48 &     76.09 &           0.053 & 15.02\% &    0.05 \\
\redsize Telefonaktiebolaget LM Ericsson ({\tiny ERICb.ST}) &          373.20 &     78.41 &           0.054 & 15.20\% &    0.05 \\
\redsize Hennes \& Mauritz AB ({\tiny HMb.ST})               &          361.66 &    232.89 &           0.112 & 11.37\% &    0.10\\
\redsize Atlas Copco AB ({\tiny ATCOa.ST})                  &          329.94 &    175.19 &           0.110 & 16.13\% &    0.10 \\
\redsize Swedbank AB ({\tiny SWEDa.ST})                     &          313.18 &    151.97 &           0.108 & 15.29\% &    0.10 \\
\redsize Sandvik AB ({\tiny SAND.ST})                       &          296.09 &     90.88 &           0.067 & 17.01\% &    0.05 \\
\redsize SKF AB ({\tiny SKFb.ST})                           &          255.99 &    161.11 &           0.112 & 16.47\% &    0.10 \\
\redsize Skandinaviska Enskilda Banken AB ({\tiny SEBa.ST}) &          221.23 &     66.85 &           0.053 & 15.56\% &    0.05 \\
\redsize Nokia OYJ ({\tiny NOKI.ST})                        &          209.77 &     28.84 &           0.019 & 36.89\% &    0.01 \\
\redsize Telia Co AB ({\tiny TLSN.ST})                      &          207.09 &     45.14 &           0.014 & 10.13\% &    0.01 \\
\redsize ABB Ltd ({\tiny ABB.ST})                           &          179.51 &    144.35 &           0.108 & 11.89\% &    0.10 \\
\redsize AstraZeneca PLC ({\tiny AZN.ST})                   &          168.06 &    318.57 &           0.127 & 12.09\% &    0.10 \\
\bottomrule
\end{tabular}  }
  \caption{Statistics of the 13 studied stocks. Values and prices are in Swedish Krona. The GK-volatility is yearly estimated.
  The table is sorted by the average daily traded value over 180 trading days.}
  \label{tab:stat:des:stocks:market}
\end{table}

Table \ref{tab:stat:des:stocks:market} discloses descriptive statistics on the considered stocks in the database.
Stocks are ranked by the average daily traded value (in unites of $10^6$ of the local currency, the Swedish krona), which can be considered as an indicator of liquidity. We also included in Table \ref{tab:stat:des:stocks:market} the average price during the study period, since European exchanges apply dynamic tick size schedules. The lower the average stock price, the lower is the tick size (see Chapter 1, Section 3 of \cite{citeulike:12047995}). The \emph{Minimum tick size} is the smallest tick size which was applied to the stock price during our study period. If the price changes are large enough, different tick sizes could have been applied during the study period, therefore we also added the yearly estimated Garman and Klass volatility to the table (see \cite{gklass80}).
Last but not least the \emph{average bid-ask spread} has to be compared with the tick size: for all these stocks the bid-ask spread lays between one and two ticks. All this stocks are therefore liquid and large tick stocks.

\begin{table}[ht!]
  \centering
{\small
\begin{tabular}{lrrrrrrr}
\toprule
Code &  Global &   HF MM & Instit. & HF Prop. &   Trades &  Pct. & Pct. LOB \\
    &    Banks &         & Brokers &          &    Count &  Ident. &  matched           \\
\midrule
VOLVb.ST &              56.9\% &       17.1\% &                 10.6\% &          15.3\% & 927,467 &    76.7\% &      97.4\% \\
NDA.ST   &              60.7\% &       10.6\% &                  9.9\% &          18.7\% & 694,509 &    76.8\% &      97.4\% \\
ERICb.ST &              57.8\% &       17.6\% &                  7.7\% &          16.9\% & 811,931 &    81.0\% &      97.2\% \\
HMb.ST   &              58.5\% &       16.0\% &                  8.9\% &          16.6\% & 716,644 &    76.8\% &      97.8\% \\
ATCOa.ST &              58.2\% &       13.7\% &                 10.5\% &          17.6\% & 677,981 &    79.1\% &      98.0\% \\
SWEDa.ST &              61.2\% &       12.2\% &                  9.5\% &          17.2\% & 600,655 &    74.6\% &      97.7\% \\
SAND.ST  &              61.0\% &       15.2\% &                 10.4\% &          13.4\% & 701,961 &    77.4\% &      96.9\% \\
SKFb.ST  &              60.9\% &       13.8\% &                 10.4\% &          14.9\% & 587,088 &    77.1\% &      97.0\% \\
SEBa.ST  &              61.5\% &       12.1\% &                  8.8\% &          17.7\% & 515,743 &    75.8\% &      97.8\% \\
NOKI.ST  &              54.5\% &        8.1\% &                  8.9\% &          28.5\% & 710,173 &    79.6\% &      99.2\% \\
TLSN.ST  &              61.2\% &       10.0\% &                 10.6\% &          18.2\% & 548,602 &    68.9\% &      97.8\% \\
ABB.ST   &              50.1\% &       15.6\% &                  5.2\% &          29.2\% & 359,067 &    86.2\% &      98.1\% \\
AZN.ST   &              51.4\% &       12.8\% &                  9.0\% &          26.8\% & 411,118 &    89.6\% &      98.8\% \\
\midrule
Average      &              58.3\% &       13.6\% &                  9.4\% &          18.7\% & --- &    77.7\% &      --- \\
\bottomrule
\end{tabular}  }
  \caption{Statistics on labelled trades involving each kind of market participant.
    \emph{Trades Count} is the sum of trades involving at least one labeled participant. 
    \emph{Pct. Ident.} represents the percentage of trades involving at least one participant of out of the four types that we focus on.
    \emph{Pct. LOB matched} is the percentage of trades for which we found a matching quote in our LOB database. The average at the bottom line is calculated over all identified trades.}
  \label{tab:stat:des:stocks:participants}
\end{table}

The NASDAQ OMX database contains the identity of the buyer and the seller \emph{from the viewpoint of the exchange}, that is, the members of the exchange who made the transactions. Asset managers for example, are not direct members of the exchange. On the other hand, brokers, banks and some other specific market participants are members.
We classify the market members into four types (for more details see Appendix \ref{sec:compap}):
\begin{itemize}
\item Global investment banks (GIB);
\item Institutional brokers (IB);
\item High frequency market makers (HFMM);
\item High frequency proprietary traders (HFPT).
\end{itemize}
Table \ref{tab:stat:des:stocks:participants} gives some plain statistics about the number of trades on each stock of our database involving these types of participants. Keep in mind that  the database covers 180 trading days. 
It can be read on the last line that, on average, Global investment banks are involved in 58\% of the trades while High Frequency Traders are involved into 32\% of them, the remaining 10\% involve institutional brokers.
The percentage of identified participant is on average 78\%, that is, 22\% of the trades took place between two participants which we are not associated with any of our four classes (GIB, IB, HFMM, HFPT). Moreover, we had to filter around 2\% of the trades (see last column) because of some cases where we could not match limit orderbook records with the observed transactions.

We expect institutional brokers to execute orders for clients without taking additional risks (i.e. act as ``pure agency brokers''). Such brokers often have medium size clients and local asset managers. They do not spend of lot of resources such as technology or quantitative analysts to study the microstructure and react fast to microscopic events.

Global Investment Banks can take risks at least on a fraction of their order flow. Most of them already had proprietary trading desks and high frequency trading activities in 2013 (i.e. during the recording of the data). They usually have large international clients and have the capability to react to changes in the state of the order-book. 

High frequency market makers are providing liquidity on both sides of the order book. They have a very good knowledge on market microstructure. As market makers, we expect them to focus on adverse selection, and not to keep large inventories.
On the other hand, high frequency proprietary traders take their own risks in order to earn money, while taking profit of their knowledge of the order book dynamics.

% IMSERT TABLE
% \rowcolor{Gray}
\begin{table}[ht!]
  \centering
  \begin{tabular}{llrrr} 
\toprule
Participant Class &Trade & Average & Average & Pct. \\%\hline
                  &Type  & imbalance & Number &\\\midrule
\rowcolor{Gray}Global Banks & Limit &             -0.41 & 103,418 &           48.2\% \\
\rowcolor{Gray}             & Market &              0.56 & 111,082 &           51.8\% \\
HF MM & Limit &             -0.31 &  30,747 &           73.0\% \\
             & Market &              0.62 &  11,818 &           27.0\% \\
\rowcolor{Gray}HF Prop. Traders & Limit &             -0.37 &  28,763 &           47.2\% \\
\rowcolor{Gray}             & Market &              0.63 &  31,858 &           52.8\% \\
Instit. Brokers & Limit &             -0.56 &   9,984 &           33.6\% \\
             & Market &              0.33 &  19,505 &           66.4\% \\
\bottomrule
  \end{tabular}
  \caption{Descriptive statistics of market participants on an ``average stock''. All the trades are normalized as if all orders were buy orders. The imbalance is positive when its sign is in the direction of the trade.}
  \label{tab:descstat}
\end{table}

% All trades from January to September 2013 on AstraZeneca (it 362,728 trades)

The data in Table \ref{tab:descstat} is compatible with our prior knowledge on the different classes of traders: 
\begin{itemize}
\item HFMM trade far more with limit orders (73\%), than with market orders;
\item IB use more market orders than limit orders;
\item on average, HFPT and GIB have balanced order flows.
\end{itemize}
Moreover, high frequency participants (HFMM and HFPT) both use market orders to consume liquidity on the \emph{weak size of the book} (i.e. buying when the imbalance is on average 0.60 and selling when it is on average -0.60), and provide liquidity when the imbalance is less intense than -0.5. The latter observation can be compatible with HF participant contributing to stabilize the price with their limit orders.

\begin{figure}[ht!]
  \centering
  \includegraphics[width=\linewidth]{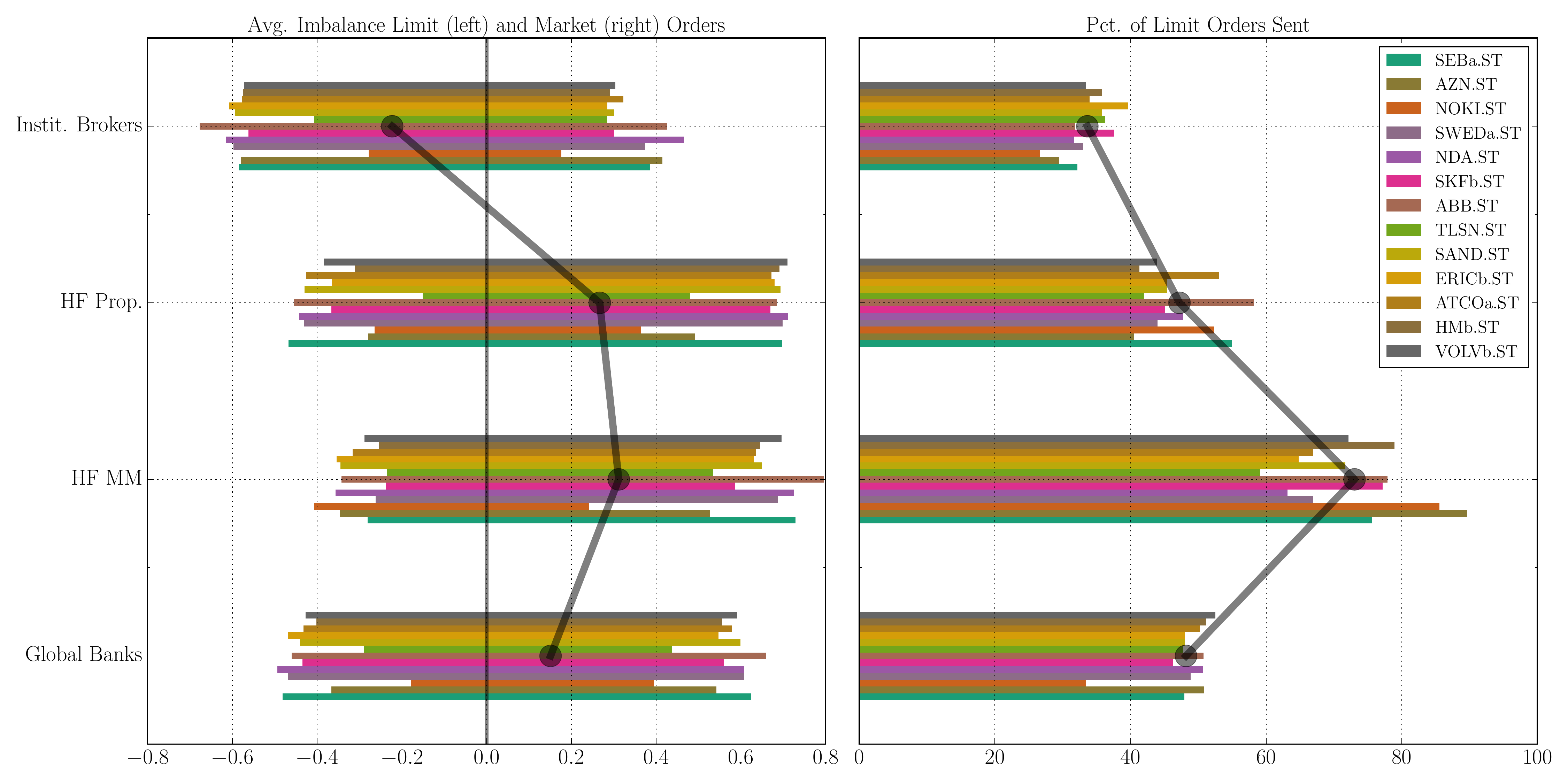}
  \caption{Use of limit and market orders and state of the imbalance before a trade, for each type of market participant.
   On the left panel: average imbalance just before a limit order (left part, negative), and average imbalance just before a market order (right part, positive). The dark line with the large dots represents the average over all trades for all stocks.
    On the right panel: percentage of trades with limit orders out of all orders. 
    The dark line is an average over all stocks.}
  \label{fig:imb:otype}
\end{figure}

These numbers are only averages, in Figure \ref{fig:imb:otype} we give their dispersion across our 13 stocks.
It can be seen in Figure \ref{fig:imb:otype} that the asymmetry between HFPT and IB is observed for all stocks (see left panel).
Moreover, the left panel suggests that high frequency participants use market orders and limit orders when the imbalance is in their favour.
%Moreover, the average trade size column shows an expected ranking: GIB first, then HFPP, followed by Institutional Brokers and HFMM.
% The average imbalance column in Table \ref{tab:descstat} is addressed in the next section.

\subsection{The Imbalance Signal}
\label{sec:imb:signal}

The order book imbalance has been identified as one of the main drivers of liquidity dynamics. It plays an important role  in order-book models and more specifically it drives the rate of insertions and cancellations of limit orders near the mid price (see \cite{aber16book, citeulike:12810809}). As an illustration of the theoretical results of this paper, we document here the \emph{imbalance signal} and its use by different types of participants. This signal is computed by using the quantity of the best bid $Q_{B}$ and the best ask $Q_A$ of the order book, 
$$\mbox{Imb}(\tau)=\frac{Q_B(\tau) - Q_A(\tau)}{Q_B(\tau) + Q_A(\tau)},$$
just before the occurrence of a transaction at time $\tau^+$.
%Note AstraZeneca is a ``\emph{medium tick stock}'' in the sense its bid-ask spread is on average 1.27 times the tick. 
Note that our 13 stocks are considered as ``\emph{large tick stocks}'' except from Sandvik AB (SAND.ST) and Telia Co AB (TLSN.ST) for which the average bid-ask spread is greater than 1.4 times the tick size.
This means that the liquidity at the best bid and ask gives a substantial information on the price pressure (see \cite{citeulike:13800065} for details about the role of the tick size in liquidity formation). For smaller tick stocks, several price levels need to be aggregated in order to obtain the same level of prediction for future price moves.

% INSERT FIGURE
\begin{figure}[ht!]
  \centering
  \includegraphics[width = 10cm, trim={0 0 25cm 0},clip]{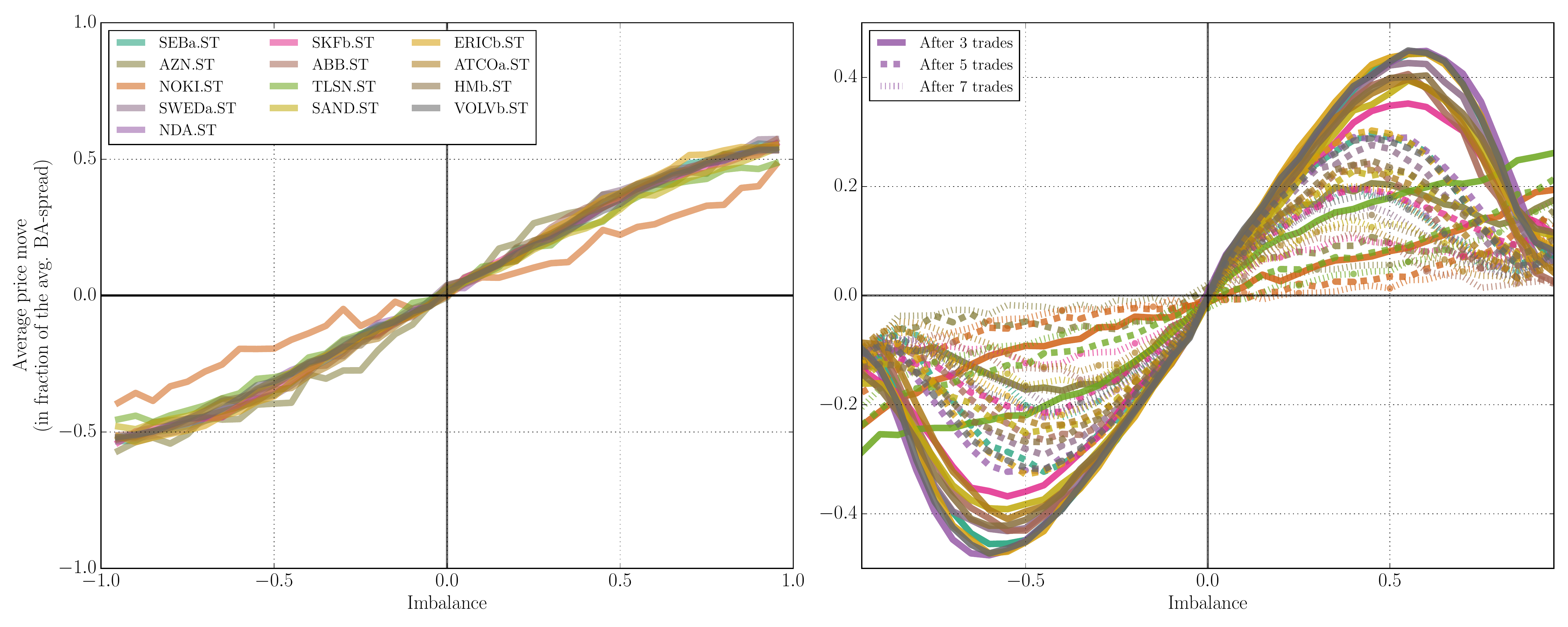}
  \caption{Predictive power of the imbalance: the average price move for the next 10 trades ($y$-axis), as a function of the current imbalance ($x$-axis).}
  \label{fig:preimb}
\end{figure}

\begin{figure}[ht!]
  \centering
  \includegraphics[width=10cm ]{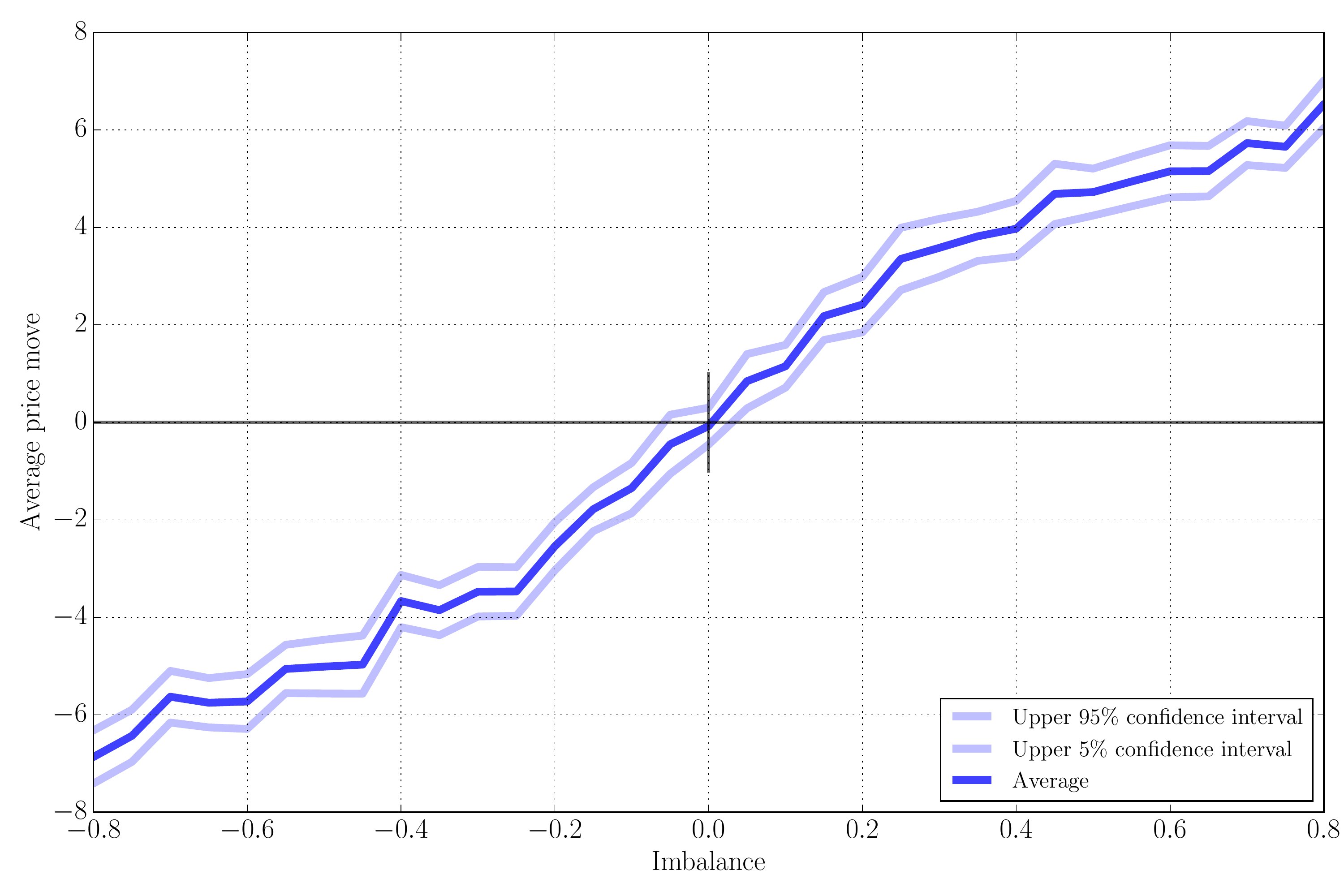}
  \caption{Predictive power of the imbalance for the AstraZeneca stock: the average price move for the next 10 trades ($y$-axis), as a function of the current imbalance ($x$-axis), with confidence levels of upper and lower $5$ precent.}
  \label{fig:preimb2}
\end{figure}

\begin{figure}[ht!]
  \centering
  \includegraphics[width = 12cm, trim={26cm 0 0 0},clip]{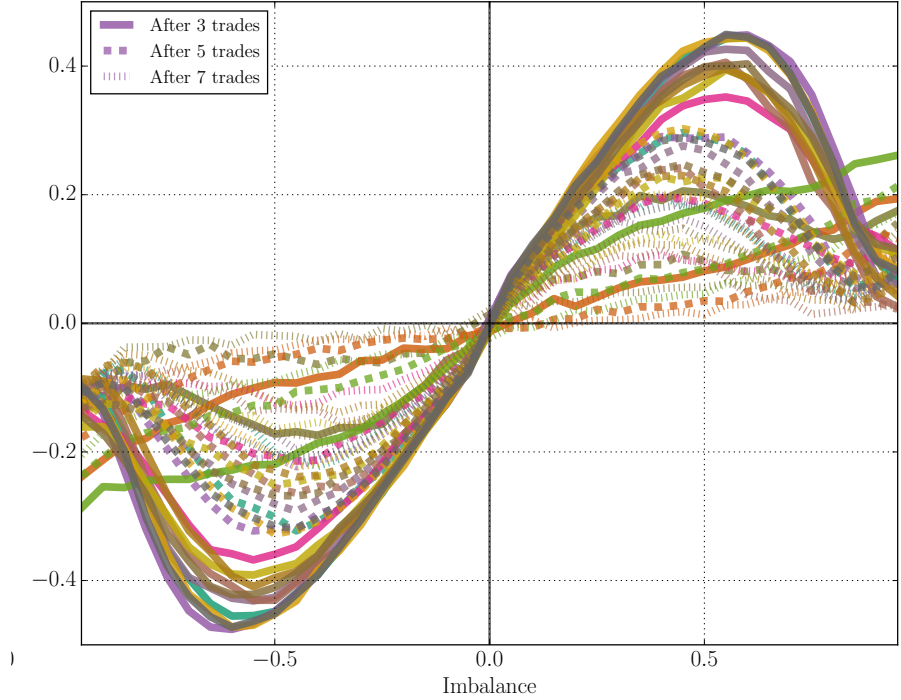}
  \caption{Mean-reversion of the imbalance: the average value of the imbalance after 3 (solid lines), 5 (dashed lines) and 7 (dotted lines) trades ($y$-axis), as a function of the current imbalance ($x$-axis). The colors of the lines represent the same stocks as in Figure \ref{fig:preimb}.}
  \label{fig:preimb3}
\end{figure}

\begin{table}[ht!]
  \centering
  {\small
\begin{tabular}{lrr|rrrrr}
\toprule
{} & d Price &  $R^2$ & Imb. 3t & Imb. 5t & Imb. 7t & Imb. 10t & Imb. 100t\\
\midrule
VOLVb.ST & 0.58 & 0.16 &       0.91 &       0.72 &       0.49 &        0.26 &         0.03 \\
NDA.ST   & 0.58 & 0.16&       0.90 &       0.71 &       0.51 &        0.30 &         0.04 \\
ERICb.ST & 0.62 & 0.15 &       0.93 &       0.74 &       0.53 &        0.30 &         0.03 \\
HMb.ST   & 0.59 & 0.08 &       0.84 &       0.62 &       0.41 &        0.21 &         0.02 \\
ATCOa.ST & 0.60 & 0.13 &       0.85 &       0.58 &       0.34 &        0.13 &         0.02 \\
SWEDa.ST & 0.62 & 0.14 &       0.87 &       0.67 &       0.45 &        0.23 &         0.02 \\
SAND.ST  & 0.56 & 0.15&       0.81 &       0.57 &       0.37 &        0.20 &         0.03 \\
SKFb.ST  & 0.59 & 0.13&       0.76 &       0.49 &       0.28 &        0.13 &         0.01 \\
SEBa.ST  & 0.61 & 0.15 &       0.91 &       0.73 &       0.51 &        0.28 &         0.03 \\
NOKI.ST  & 0.41 & 0.01 &       0.18 &       0.08 &       0.05 &        0.03 &         0.00 \\
TLSN.ST  & 0.54 & 0.04 &       0.43 &       0.22 &       0.13 &        0.08 &         0.02 \\
ABB.ST   & 0.59 & 0.11&       0.86 &       0.61 &       0.33 &        0.11 &         0.03 \\
AZN.ST   & 0.64 & 0.04 &       0.47 &       0.20 &       0.09 &        0.05 &         0.02 \\
\bottomrule
\end{tabular}}    
  \caption{Results of linear regressions involving the imbalance.
  The first column is the result of a regression of the price move after 10 trades given the imbalance immediately before the first of these trades. This can also be shown in the slope of Figure \ref{fig:preimb}. The p-value is very close to zero for all stocks, meaning they are highly significant. The $R^2$ varies between 1\% (Nokia) to 16\% (Volvo AB and Nordea Bank AB).
Other columns are the result of the regression of future imbalance (respectively after 3, 5, 7, 10 and 100 trades) with respect to the imbalance immediately before the first of these trades, given that the imbalance is between -0.5 and 0.5. This regression corresponds to the slopes at the center of Figure \ref{fig:preimb3}. All p-values are significant at more than 99.99\%.}
  \label{tab:reg:imb}
\end{table}

In order to demonstrate the predictive power of the imbalance, we consider the average mid price move after 10 trades %\footnote{On our data, there is on average one trade every PP seconds.} 
as a function of the current imbalance (see Figure \ref{fig:preimb}).
Table \ref{tab:reg:imb} gives data which is associated to these curves.
The column ``\emph{d Price}'' shows the price change re-normalized by the average bid-ask spread on each stock after 10 trades. This price move is on average close to 0.6 times the imbalance just before the first of these trades;
%\item The other columns show that the correlation between the imbalance now and the imbalance after $n$ trades decays to zero with $n$. This slope corresponds to the imbalance after $n$ trades given the current imbalance is between -0.5 and 0.5, to capture the linear effect at the center of Figures \ref{fig:preimb} and \ref{fig:preimb2}. When the imbalance is more intense than 0.5, it is very likely the current best bid and ask queues will be fully depleted, and hence the ``new'' imbalance is less related to the current one. This effect probably explains the non linearity of the solid curves.
%Two stocks (Nokia and Telia) do not exhibit such a pattern. At the light of Table~\ref{tab:stat:des:stocks:market}, it may be because these two stocks are ``smaller tick'' stocks than the others.
%\end{itemize}
%Keep in mind that we are not performing a detailed study of the imbalance signal. The charts and tables in this section are mainly informative and intend to justify our theoretical models.

\paragraph{Mean-reversion of the imbalance.}
Figure \ref{fig:preimb3} on the right shows the average value of the imbalance after $\Delta T =3, 5$ and $7$ trades as a function of its current value. The colors of the curves represent the same stocks as in Figure \ref{fig:preimb}.
The decreasing slopes around $\mbox{Imb}(t)=0$ are underlined by the last columns $4-6$ of Table \ref{tab:reg:imb}. It demonstrates the mean reverting property of the imbalance.
We will not comment too much on the decreasing slopes for large imbalance values. We will just mention that strong imbalance may imply on a future price change, which in turn, can create a depletion of the ``weak side of the order-book'' (in the sense of \cite{citeulike:12335801}). This phenomenon may cause an inversion of the imbalance, since the queue in second best price level of the order book, which is now ``promoted'' to be the first level, could be large. See \cite{citeulike:12810809} for details about queues dynamics in order-books.

\begin{figure}[ht!]
  \centering
  \includegraphics[width=1.\linewidth]{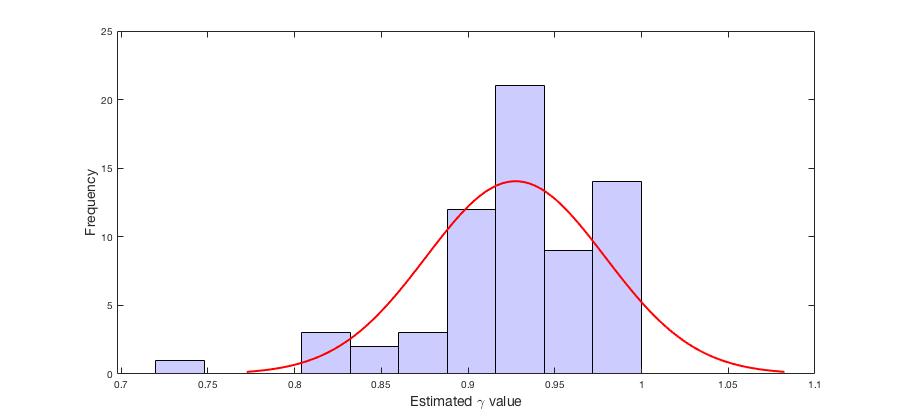}
  \caption{Histogram for the estimators of $\gamma$.}
  \label{fig:imb:gamma}
\end{figure}

To approximately fit Ornstein-Uhlenbeck (OU) dynamics to the imbalance data, we will use ``trade time'' instead of ``calendar time'' (i.e. seconds), in order to compensate on different frequencies of trading for each of our 13 stocks (see columns $dt$ on Table~\ref{tab:imb:sigma}  in the Appendix). This implies a discrete version of an OU, 
$$I_{n+\Delta n} - I_n= -\gamma \cdot I_n \Delta n + \sigma \sqrt{\Delta n} \cdot \xi_{n+\Delta n},$$
where $\Delta n$ is the number of trades ahead you look at, $\gamma$ is the speed of mean-reversion parameter and $\sigma$ is the standard deviation of the innovation $\xi_{n+\Delta n}$. The linear regressions on the last columns of Table~\ref{tab:reg:imb} are following the model
$$I_{n+\Delta n} = a_{\Delta n} \cdot I_n + \tilde\sigma_{\Delta n}\cdot\epsilon_{n+\Delta n}.$$

This give a simple identification which leads to the following estimators of $\gamma$ and $\sigma$:
$$\hat\gamma \simeq \frac{1 - a_{\Delta n}}{\Delta n},\quad \hat\sigma \simeq \frac{\tilde\sigma_{\Delta n}}{\sqrt{\Delta n}}.$$
Figure~\ref{fig:imb:gamma} shows the frequencies of values of $\hat\gamma$ for the 13 stocks over the scales $\Delta n = 3,5,7,100$. Table~\ref{tab:imb:gamma} gives the associated values of these $\hat\gamma$'s. In Table~\ref{tab:imb:sigma} different estimates of $\hat\sigma$ are given.

\paragraph{Some numerical values of the model parameters.}\label{sec:emp:param:values}
At a time scale of 35 seconds or 7 trades, $\gamma$ should be taken close to 0.92 and $\sigma$ close to 0.22.
We also provide an estimator for the instantaneous market impact $\kappa$ using the empirical average of the mid-price move\footnote{The mid-price is the middle of the best bid and best ask prices.} after a trade times the sign of the trade.
Table~\ref{tab:imb:sigma} in the Appendix shows that the average value of $\kappa$ divided by the average bid-ask spread is close to 0.1.
%Moreover, we had to normalize the price by the bid-ask spread of each stock to obtain the aligned straight lines of Figure~\ref{fig:preimb} (left).
%For numerical applications we can assume a bid-ask spread
%In both cases we observe a decrease with the scale, probably because the innovations $\xi_{n+\Delta n}$ exhibit some mean reversion properties.

We summarise the main findings of this section: 
\begin{itemize}
\item[\textbf{(i)}] the imbalance can be considered as a liquidity-driven short term signal,
\item[\textbf{(ii)}] this signal has mean-reverting properties,
\item [\textbf{(ii)}] Narket participants, especially high frequency traders, take the imbalance into account while trading (see Table \ref{tab:descstat} and Figure~\ref{fig:imb:otype}).

%%\item some market participants adjust their short term behaviour according to the value of $\mbox{Imb}(\tau)$. 
\end{itemize}

\subsection{Use of signals by market participants} \label{sec-use-sig}

As previously mentioned, we expect HF Proprietary Traders, HF Market Makers and Global Investment Banks to pay more attention to order-book dynamics than Institutional Brokers.
However, as market makers, HFMM are expected to earn money by buying and selling when the mid price does not change much (relying on the \emph{bid-ask bounce}). On the other hand, HFPT are typically alternating between intensive buy and sell phases which are based on price moves. 

% \begin{figure}[ht!]
%   \centering
%   \includegraphics[width=.8\linewidth]{p02_AZN_avg_imb_both_01_mm.pdf}
%   \caption{The average imbalance before a trade, using a limit buy order (blue) or a market buy order (red).}
%   \label{fig:imbbefore}
% \end{figure}

Our expectations are met in % Figure \ref{fig:imbbefore}
Table \ref{tab:descstat}, where the average imbalance just before a trade is shown for each type of market participant. All the trades in this table are normalized as if all orders were buy orders. The imbalance is positive when its sign is in the direction of the trade, and negative if it is in an apposite direction.

We notice the following behaviour:
\begin{itemize}
\item when the transaction is obtained via a market order, the market participant had the opportunity to observe the imbalance before consuming liquidity.
\item when the transaction is obtained via a limit order, fast participants have the opportunity to cancel their  orders to prevent an execution and potential adverse selection.
\end{itemize}

Table \ref{tab:descstat} % Figure \ref{fig:imbbefore} 
underlines that HF participants and GBI make ``better choices'' on trading according to the market imbalance.
Institutional Brokers seems to be the less ``imbalance aware'' when they decide to trade. This could be explained either by the fact that they invest less in microstructure research, quantitative modelling and automated trading; or either because they have less freedom to be opportunistic. Since they act as pure agency brokers, they do not have the choice to retain clients orders, and it could prevent them from waiting for the best imbalance to trade. 
%These charts are not dynamics, they do not take into account the 

\paragraph{Strategic behaviour.}
Once we suspect that some participants take into account the imbalance in their trading decisions; we can look for a relation between the trading rate and the corresponding imbalance for each type of participant.
This is motivated by the optimal trading frameworks of previous sections, where we used the trading rate as a control.
\begin{figure}[ht!]
  \centering
  \includegraphics[width=\linewidth]{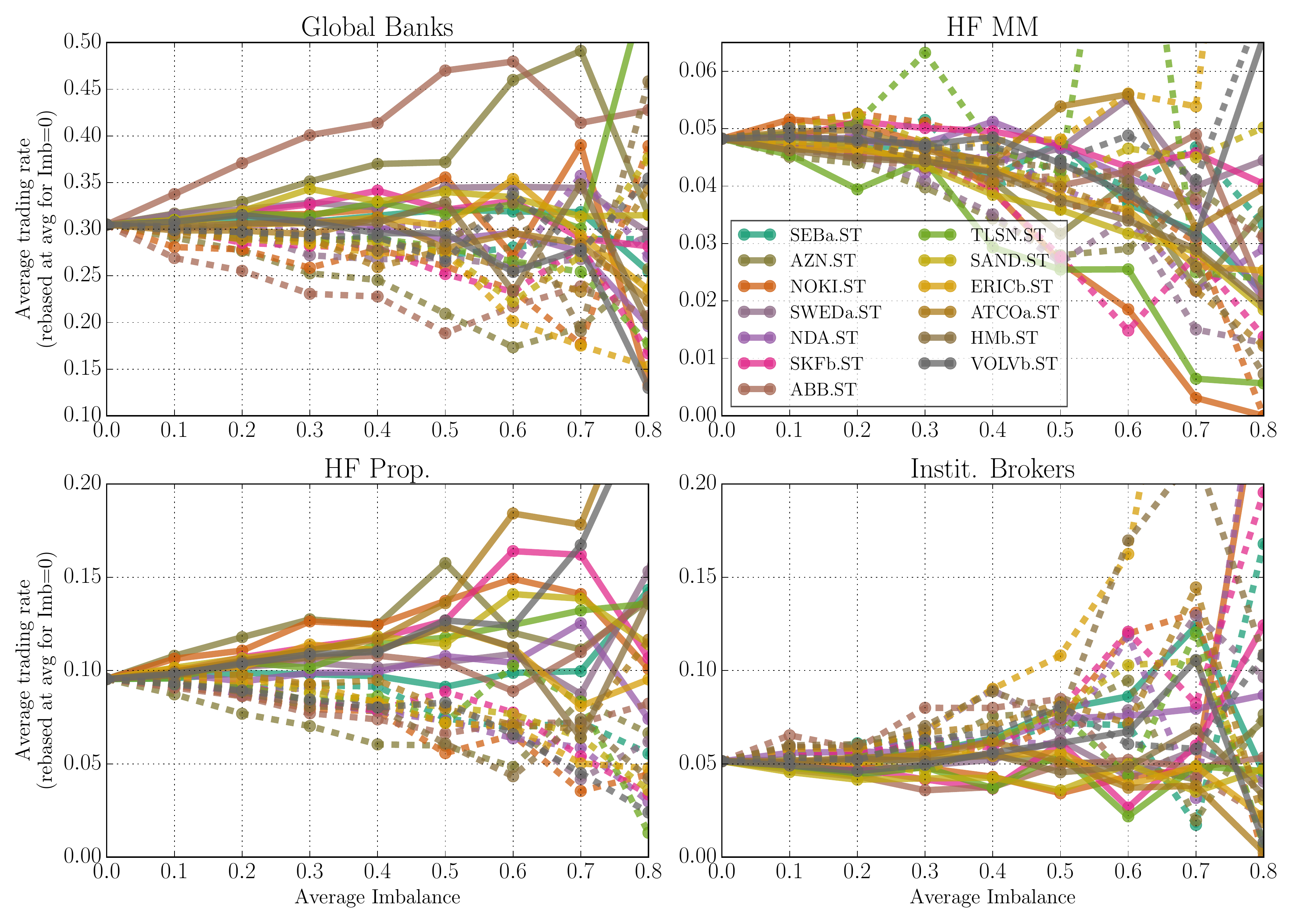}
  \caption{Renormalized average trading rate in the direction of the imbalance ${\hat r}_+$ (solid line) and in the opposite direction ${\hat r}_-$ (dotted line), during 10 consecutive minutes, for each type of participant.}
  \label{fig:imbrate}
\end{figure}

In order to learn more about the relation between the imbalance signal and the trading speed, we compute the \emph{imbalance-conditioned trading rate} $R_+$ and $R_-$ for each type of market participant, during all consecutive intervals of 10 minutes from January 2013 to September 2013 (within the trading hours, i.e. 9h00 to 17h30). Note that in the following analysis the signal, time and trading quantities are discrete. 

\begin{definition}[Imbalance conditioned trading quantities]
  The {Imbalance conditioned trading rate} of market participants of type $\calP$ during the time interval $\calT$
  are given by 
  $$R_\pm(\calT,\calP\;|\;\iota) = \frac{1}{N(\calT,\calP,\iota)} %
  \sum_{t\in\calT} \bar\delta_{\varepsilon(t)\cdot\textrm{sign}({\rm Imb}(t))}(\pm\iota) \cdot A_t \cdot \bar \delta_\calP(t)\cdot\bar \delta_{|Imb(t)|}(\iota),$$
where  \begin{itemize}
  \item $\varepsilon(t)$ is the sign of the trade at time $t$.
  \item $\bar \delta_{\varepsilon(t)\cdot\textrm{sign}({\rm Imb}(t))}(\pm1)$ is $1$, if at time $t$ the imbalance sign times the sign of the trade is equal to $\pm1$, and $0$ otherwise.
  \item $A_t$ is the traded amount of the trade at time $t$.
  \item $\bar \delta_\calP(t)$ is $1$ if the trade at time $t$ involved a participant of type $\calP$, and $0$ otherwise. 
  \item $\bar \delta_{|Imb(t)|}(\iota)$ is $1$ if the absolute value of the imbalance at time $t$ equals to $\iota$, and $0$ otherwise.
  \item $N(\calT,\calP,\iota)$ is the of number of trades involving participant $\calP$ in $\calT$ when the imbalance is equals to $\iota$.
  \end{itemize}
 Qualitatively, $R_{\pm}$ have the following interception,
  \begin{itemize}
  \item $R_+(\calT,\calP\;|\;\iota)$ is an estimate of the amount traded \emph{in the direction of the imbalance}, when the absolute value of the imbalance is $\iota$, by participants of type $\calP$ during the time interval interval $\calT$,
  \item $R_-(\calT,\calP\;|\;\iota)$ is an estimate of the amount traded \emph{in the opposite direction of the imbalance}, when the absolute value of the imbalance is $\iota$, by participants of type $\calP$ during interval $\calT$.
  \end{itemize}
\end{definition}
In order to get the trading imbalance conditioned rates we renormalize $R_{\pm}$ by $A(\calT \;|\; \iota)$ the traded amount during the interval $\calT$ given the imbalance is $\iota$:
$$A(\calT \;|\; \iota)=\sum_\calP R_+(\calT,\calP\;|\;\iota) + R_-(\calT,\calP\;|\;\iota).$$
Then ${R_+(\calT,\calP\;|\;\iota)}$ divided by ${A(\calT\;|\; \iota)}$ is an estimate of \emph{the probability that a stock is traded by a participant of type $\calP$ during interval $\calT$ in the direction of the imbalance, given that the imbalance is $\iota$}. 
${R_-(\calT,\calP\;|\;\iota)}$ divided by ${A(\calT \;|\; \iota)}$ is an estimate to the probability that a stock is traded by a participant of type $\calP$ during interval $\calT$, in an opposite direction of the imbalance, given the imbalance is $\iota$.

Let $N_\calT$ be the number of ten minutes intervals in our data base. We define
$$r_+(\calP\;|\;\iota)=\frac{1}{N_\calT} \sum_\calT \frac{R_+(\calT,\calP\;|\;\iota)}{A(\calT \;|\; \iota)},\quad %
r_-(\calP\;|\;\iota)=\frac{1}{N_\calT} \sum_\calT \frac{R_-(\calT,\calP\;|\;\iota)}{A(\calT \;|\; \iota)},$$
which are unbiased estimators for the probability that a participant of type $\calP$ trades in the direction (respectively, opposite direction) of the imbalance, given that the absolute value of the imbalance is $\iota$.

To be able to put all the stocks on the same graph, we draw $${\hat r}^k_\pm(\calP\;|\;\iota) = r^k_\pm(\calP\;|\;\iota)/{\bar r}_\pm(\calP\;|\;0)$$ in Figure \ref{fig:imbrate}. Here ${\bar r}_\pm(\calP\;|\;0)$ is the average of $r^k_\pm(\calP\;|\;0)$ over all stocks $k$.

Figure \ref{fig:imbrate} shows the variations of ${\hat r}_+$ (the relative speed of trading in the direction to the imbalance, in solid lines) and ${\hat r}_-$ (the relative speed of trading in the opposite direction to the imbalance, in dashed lines) with respect to the imbalance $\iota$ before the trade, for each type of market participant and for each stock. From this graph we observe the following:
\begin{itemize}
\item High Frequency Market Makers:
  we observe that the higher the imbalance in the orderbook, the less they trade. 
  This effect does not seem to be related to the direction of their trades. 
  It corresponds to an expected behaviour from market makers.
\item High Frequency Proprietary Traders:
  the higher the imbalance, the more they trade in the similar direction, and the less they trade in the opposite direction.
\item Institutional Brokers do not seem to be influenced by the imbalance. Additional data analysis shows that they they trade more with limit orders when the imbalance is intense, this  may derive the price to move in the opposite direction.
\item The behaviour of Global Banks seems to be influenced by the imbalance for part of the stocks in our sample.  
\end{itemize}

\paragraph{Towards a theory for the strategic use of signals.}
The analysis in this section suggests that some market participants are using liquidity-driven signals in their trading strategies. The liquidity imbalance, computed from the best bid and ask prices of the order-book for medium and large tick stocks, appears to be a good candidate. Moreover, its dynamics exhibit mean-reverting properties.

The theory developed in Sections \ref{sec-res} and \ref{Sec-exmp} can be regarded as a tentative framework to model the behaviour the following participants. 
Global investment banks who execute large orders, seem to be a typical example for participants who adopt the type of strategies that we model.  
HFPT who are combining slow signals (which may be considered as execution of large orders) along with fast signals, could also use our framework.
We could moreover hope that thanks to the availability of such frameworks, Institutional Brokers could optimize their trading, and to increase the profits for more final investors. 

\section{Proofs}
\subsection{Proofs of Theorems \ref{thm-uniq}, \ref{thm-exs} and Corollary \ref{cor-cost} } \label{Sec-proofs}
The proofs of Theorems \ref{thm-uniq} and \ref{thm-exs} use ideas from the proofs of Proposition 2.9 and Theorem 2.11 in \cite{GSS}.
\paragraph{Proof of Theorem \ref{thm-uniq}}  
Let $x\geq 0$. For any $X\in \Xi(x)$ define 
\be \label{gen-cost} 
C(X):=C_{1}(X)+C_{2}(X) +K(X), 
\ee
where
\bn
C_{1}(X) &=& \frac{1}{2}\int \int G(|t-s|)\,dX_s\,dX_t,  \\
C_{2}(X) &=& \phi\int_{0}^{T}X_{s}^{2}ds, \\
K(X)&=&\int \int_{0}^{t}  E_{\iota}[I_{s}]\,ds\,dX_t .
\en
Note that $C(x)$ is the cost functional in (\ref{ex-cost2}). 

Since $G$ is strictly positive definite we have for any $X\in \Xi(x)$, 
\be \label{pos-c1} 
C_{1}(X)> 0. 
\ee
$C_{2}(\cdot)$ is quadratic in $X$ and therefore we have 
\be \label{nneg-c2}
C_{2}(X) \geq 0. 
\ee
Let $X,Y \in \Xi(x)$. We define the following cross functionals, 
\bn
C_{1}(X,Y) &=&  \frac{1}{2}\int \int G(|t-s|)\,dX_s\,dY_t , \\
C_{2}(X,Y)&=&\phi \int_{0}^{T}X_{s}Y_{s}ds.
\en 
Note that 
\bn
C_{i}(X,Y)  = C_{i}(Y,X), \quad \textrm{for } i=1,2, 
\en
and
\be \label{cx-y}
C_{i}( X- Y) =C_{i}(X)+C_{i}(Y) -2C_{i}(X,Y), \quad \textrm{ for  } \ i=1,2. 
\ee
From (\ref{pos-c1}) it follows that $C_{1}(X-Y)>0$ and together with (\ref{cx-y}) we get 
\bn
C_{1}\Big(\frac{1}{2} X+ \frac{1}{2}Y\Big) &=& \frac{1}{4}C_{1}(X)+\frac{1}{4}C_{1}(Y) +\frac{1}{2}C_{1}(X,Y) \\
&<&\frac{1}{2}C_{1}(X)+\frac{1}{2}C_{1}(Y).
\en
Repeating the same steps, using (\ref{nneg-c2}) instead of (\ref{pos-c1}) we get 
\bn
C_{2}\Big(\frac{1}{2} X+ \frac{1}{2}Y\Big) &\leq &\frac{1}{2}C_{2}(X)+\frac{1}{2}C_{2}(Y).
\en
Since $K(X)$ is linear in $X$ we have 
\bn
K\Big(\frac{1}{2} X+\frac{1}{2} Y\Big) = \frac{1}{2} K\big( X\big) + \frac{1}{2} K\big( Y\big).
\en
 From (\ref{gen-cost}) it follows that
\bn
C\Big(\frac{1}{2} X+\frac{1}{2} Y\Big) < \frac{1}{2} C\big( X\big) + \frac{1}{2} C\big( Y\big).
\en
Let $\alpha \in (0,1)$. The claim that 
\bn
C\big(\alpha X+(1-\alpha) Y\big) < \alpha C\big( X\big) +(1-\alpha) C\big( Y\big), 
\en
follows from the continuity of $C(\cdot)$, by a standard extension argument. Since $C(\cdot)$ is strictly convex, we get that there exists at most one minimizer to $C(X)$ in $\Xi(x)$.

\qed
\paragraph{Proof of Theorem \ref{thm-exs}} 
First we prove that (\ref{cond-opt}) is necessary for optimality. Let $0\leq t<t_{0} \leq T$ and consider the round trip 
\bn
dY_{s} = \delta_{t_{0}}(ds) - \delta_{t}(ds). 
\en
For all $\alpha \in \re$ we have 
\bq \label{eq-c-s} 
 C_{i}(X^{*}+\alpha Y) =  C_{i}(X^{*}) + \alpha^{2}  C_{i}(Y) +2\alpha  C_{i}(X^{*},Y), \quad i=1,2, 
\eq
and 
\bq \label{eq-K} 
K(X^{*}+\alpha Y) =  K(X^{*}) + \alpha K(Y) 
\eq
Let $Z:=X^{*}+\alpha Y$, and recall that $C(Z)=C_{1}(Z)+C_{2}(Z)+K(Z)$. 
Using (\ref{eq-c-s}) and (\ref{eq-K}) we can differentiate $C(Z)$ with respect to $\alpha$ and get 
\bn
\frac{\partial C(Z)}{\partial \alpha} =K(Y)+ \sum_{i=1,2} 2\alpha C_{i}(Y) +2 C_{i}(X^{*},Y). 
\en
From optimality we have $C(X^{*}) \leq C(Z)$ and therefore we expect that
\bq \label{der}
\frac{\partial C(Z) }{\partial \alpha}\Big|_{\alpha =0}= K(Y)+2\sum_{i=1,2} C_{i}(X^{*},Y)  =0.
\eq
Note that 
\bn
 C_{1}(X^{*},Y) &=&\frac{1}{2} \int \int G(|r-s|)dX^{*}_{s}dY_{r} \\ 
 &=&\frac{1}{2} \int G(|t_{0}-s|)dX^{*}_{s}-\frac{1}{2} \int G(|t-s|)dX^{*}_{s}, 
\en
\bn
C_{2}(X^{*},Y)&=&\phi\int_{0}^{T}X^{*}_{s}Y_{s}ds\\
&=&-\phi\int_{t}^{t_{0}}X^{*}_{s}ds.
\en
and 
\bn
K(Y)&=&\int \int_{0}^{r}  E_{\iota}[I_{s}]\,ds\,dY_r  \\
&=&\int_{t}^{t_{0}}  E_{\iota}[I_{s}]\,ds.
\en
We get that (\ref{der}) is equivalent to 
\bn
&&\int G(|t_{0}-s|)dX^{*}_{s} - 2\phi\int_{0}^{t_{0}}X^{*}_{s}ds + \int_{0}^{t_{0}} E_{\iota}[I_{s}]ds\\
&&= \int G(|t_{}-s|)dX^{*}_{s}-2\phi\int_{0}^{t}X^{*}_{s}ds + \int_{0}^{t} E_{\iota}[I_{s}]ds. 
\en 
Since $t$ and $t_{0}$ were chosen arbitrarily this implies (\ref{cond-opt}). \medskip \\
Assume now that there exists $X^{*}\in \Xi(x)$ satisfying (\ref{cond-opt}), we will show that $X^{*}$ minimizes $C(\cdot)$. Let $ X $ be any other strategy in $\Xi(x)$.  Define $Z= X -X^{*}$. Then from (\ref{cond-opt}) we have 
\be \label{g1}
\aligned 
C_{1}(X^{*},Z)&=\frac{1}{2} \int \int G(|t-s|)\,dX^{*}_{s}\,dZ_{t} \\
&=\frac{1}{2} \int \Big(\lam +2\phi\int_{0}^{t}X^{*}_{s}ds - \int_{0}^{t}E_{\iota}[I_{s}]ds \Big)dZ_{t} \\
&=\frac{\lam}{2}\big(X([0,\infty))-X^{*}([0,\infty))\big)+\phi\int \int_{0}^{t}X^{*}_{s}\,ds\,dZ_{t}-\frac{1}{2} \int  \int_{0}^{t}   E_{\iota}[I_{s}]ds  \,dZ_{t}  \\
&= \phi\int \int_{0}^{t}X^{*}_{s}\,ds\,dZ_{t}-\frac{1}{2}K(Z),
\endaligned 
\ee
where we have used the fact that $X([0,\infty))=X^{*}([0,\infty))=x$ in the last equality.

From (\ref{eq-c-s}) and (\ref{g1}) we have 
\bn
C_{1}( X) &=&  C_{1}(Z+ X^{*}) \\
 &=& C_{1}(Z) +  C_{1}(X^{*}) +2 C_{1}(X^{*},Z)\\
 &=&   C_{1}(Z) +  C_{1}(X^{*}) -K(Z)+2\phi\int \int_{0}^{t}X^{*}_{s}dsdZ_{t}, 
\en
and 
\bn
  C_{2}( X) &=&  C_{2}(Z+ X^{*}) \\
 &=&  C_{2}(Z) +  C_{2}(X^{*}) +2 C_{2}(X^{*},Z) \\
 &=&  C_{2}(Z) +  C_{2}(X^{*})+2\phi\int_{0}^{T}X^{*}_{s}Z_{s}ds.
\en
From the linearity of $K(\cdot)$ we get 
\bn
K( X) &=&  K(Z) + K(X^{*}).
\en
It follows that 
\bn
C(X)&=&\sum_{i=1,2}C_{i}(X)+K(X)\\
&=&  C_{1}(X^{*}) +C_{2}(X^*)+K(X^{*})+C_{1}(Z) +C_{2}(Z) \\
&& +2\phi\int \int_{0}^{t}X^{*}_{s}\,ds\,dZ_{t} +2\phi\int_{0}^{T}X^{*}_{s}Z_{s}ds\\
&=& C(X^{*})+C_{1}(Z) +C_{2}(Z) \\
&&\quad +2\phi\int \int_{0}^{t}X^{*}_{s}\,ds\,dZ_{t} +2\phi\int_{0}^{T}X^{*}_{s}Z_{s}ds.
\en
Racal that $Z_{0}=0$ and $Z_{t}=0$ for every $t>T$, hence from integration by parts we have  
\bn
0= \int \int_{0}^{t}X^{*}_{s}\,ds\,dZ_{t} +\int_{0}^{T}X^{*}_{t}Z_{t}\,dt, 
\en
and since for $i=1,2$, $C_{i}(Z)\geq 0$, we get
\bd
C(X)\geq C(X^{*}). 
\ed
\qed
\paragraph{Proof of Corollary \ref{cor-cost}} 
From (\ref{I-OU}) it follows that $E_{\iota}[I_{t}]=\iota e^{-\gamma^{}t}$. Since $\phi=0$, (\ref{cond-opt}) reduces to
\bq \label{cond-mod}
\frac{\iota}{\gamma} (1-e^{-\gamma^{}t})+\kappa \rho \int_{0}^{T}e^{-\rho|t-s|}\,dX^{*}_{s} =\lambda. 
\eq
Moreover we have the fuel constraint, 
\bq \label{norm} 
\int_{0}^{T}dX^{*}_{t} =-x. 
\eq
Motivated by the example in Obizhaeva and Wang \cite{Ob-Wan2005}, we guess a solution of the from  
\be \label{sol-can} 
dX^{*}_{t}= A\delta_{0} + (Be^{-\gamma t}+C)dt+D\delta_{T}, 
\ee
where $\delta_{x}$ is the Dirac's delta measure at $x$ and $A,B,C,D$ are some constants.

Note that 
\bn
\kappa \rho \int_{0}^{t}e^{-\gamma s}e^{-\rho(t-s)}ds=\frac{\kappa \rho}{\rho-\gamma}\big(e^{-\gamma t}-e^{-\rho t}\big), 
\en
\bn
\kappa \rho\int_{t}^{T}e^{-\gamma s}e^{-\rho(s-t)}ds=\frac{\kappa \rho}{\rho+\gamma}\big(e^{-\gamma t}-e^{-\gamma T-\rho(T-t)}\big), 
\en
and therefore 
\bn
&&\kappa \rho\int_{0}^{T}e^{-\rho|t-s|}\,dX^{*}_{s} \\
&&=\kappa \rho e^{-\rho t}A+B\frac{\kappa \rho}{\rho-\gamma}\big(e^{-\gamma t}-e^{-\rho t}\big)+B\frac{\kappa \rho}{\rho+\gamma}\big(e^{-\gamma t}-e^{-\gamma T-\rho(T-t)}\big) \\
&&\quad+C\kappa\big(1-e^{-\rho t}\big)+C\kappa\big(1-e^{-\rho(T-t)}\big)+D\kappa \rho e^{-\rho(T-t)}.
\en
From (\ref{cond-mod}) it follows that
\bn
\lambda =2\kappa C+\frac{\iota}{\gamma},
\en
and together with (\ref{norm}) we get the following linear system, 
\bn
-\frac{\iota}{\gamma}e^{-\gamma t} +B\frac{\kappa \rho }{\rho-\gamma}e^{-\gamma t}+B\frac{\kappa \rho }{\rho+\gamma}e^{-\gamma t}&=&0, \\
A\kappa \rho  e^{-\rho t}-B\frac{\kappa \rho }{\rho-\gamma}e^{-\rho t}-C\kappa e^{-\rho t}&=&0, \\
-B\frac{\kappa \rho }{\rho+\gamma}e^{-\gamma T-\rho(T-t)}-C\kappa e^{-\rho(T-t)}+D\kappa \rho e^{-\rho(T-t)}&=&0, \\
A+\frac{B}{\gamma}\big(1-e^{-\gamma T}\big)+CT+D&=&-x.
 \en
From the first equation we can get
 \bn
 B=\iota \frac{\rho^2-\gamma^{2}}{2\kappa \rho^{2} \gamma}.
 \en
 and then 
 \begin{equation}  \label{constants}  
 \begin{aligned} 
% A&=\frac{1}{2+T\rho}\Big(\frac{\iota}{2\kappa \rho^{2}\gamma}\Big((\rho+\gamma)\big(1+T\rho+\gamma^{-1}(\rho-\gamma)(1-e^{\gamma T})\big)-(\rho-\gamma)e^{-\gamma T}\Big)-x\Big), \\
%A&=\frac{1}{2+T\rho}\Big(\frac{\iota}{2\kappa \rho^{2}\gamma}\Big((\rho+\gamma)\big(1+T\rho{\color{red}{-\gamma^{-1}(\rho-\gamma)(1-e^{-\gamma T})}}\big)-(\rho-\gamma)e^{-\gamma T}\Big)-x\Big), \\
A&=\frac{1}{2+T\rho}\left(\frac{\iota}{2\kappa \rho^{2}\gamma}\left\{(\rho+\gamma)\left(1+T\rho-\frac{\rho-\gamma}{\gamma}(1-e^{-\gamma T})\right)-(\rho-\gamma)e^{-\gamma T}\right\}-x\right), \\
C&=\rho A-\iota\frac{\rho+\gamma}{2\kappa \rho \gamma}, \\
D&=A-\frac{\iota}{2\kappa \rho^{2}\gamma}((\rho+\gamma)-(\rho-\gamma)e^{-\gamma T}). 
\end{aligned} 
 \end{equation}
 The optimal strategy is therefore 
 \bn
 X^{*}_{t}=x+\one_{\{t>0\}}A+Ct+\frac{B}{\gamma}\big(1-e^{-\gamma t}\big)+ \one_{\{t>T\}}D, 
 \en
 which is equivalent to (\ref{opt-spec}). 
\qed 

\subsection{Proofs of Propositions \ref{prop-hjb} and \ref{verification}} \label{Sec-hjb}
\paragraph {Proof of Proposition \ref{prop-hjb}.} 
The proof follows the same lines as the proof of Proposition 1 in \cite{Car-Jiam-2016}. 

Pluggin in the ansatz $V(t,\iota,c,x,p):=c+xp+v(t,x,\iota)$ we get 
\begin{align*}   
0=\partial_{t} v +\mathcal L^{I}v + \iota x-\phi x^{2}+\sup_{r}\big\{- r^{2}\kappa-r\partial_{x}v\big\}.
\end{align*} 
Optimizing over $r$ it follows that 
\be \label{r-hjb}
r^{*} =- \frac{\partial_{x}v}{2\kappa},  
\ee
and we get the following PDE
\begin{align}  \label{pde1}
\partial_{t} v+ \mathcal L^{I}v + \frac{1}{4\kappa}\big(\partial_{x}v\big)^{2} + \iota x-\phi x^{2}=0.
\end{align} 
where $v(T,x,\iota) = - \varrho x^{2}$.

As in equation (A.2) in \cite{Car-Jiam-2016}, we have a linear and quadratic $x$-terms in (\ref{pde1}) along with a quadratic terminal condition, hence we make the following ansatz on the solution: 
$$
v(t,x,\iota) = v_{0}(t,\iota) + xv_{1}(t,\iota)+x^{2}v_{2}(t,\iota). 
$$
By comparing terms with similar powers of $q$, we get the following system of PDEs, 
\bq
\label{v0}\partial_{t} v_{0}+\mathcal L^{\iota}v_{0}+ \frac{1}{4\kappa}v_{1}^{2} &=&0, \\   
\label{v1}\partial_{t} v_{1}+\mathcal L^{\iota}v_{1}+ \frac{1}{\kappa}v_{2}v_{1}+ \iota &=&0, \\
\label{v2} \partial_{t} v_{2}+\mathcal L^{\iota}v_{2}+\frac{1}{\kappa}v_{2}^{2} -\phi&=&0, 
\eq
with the terminal conditions
$$
v_{0}(T,\iota)=0, \quad v_{1}(T,\iota) =0, \quad v_{2}(T,\iota)= -\varrho. 
$$
We first find a solution to (\ref{v2}). Note that since the terminal condition is independent of $\iota$ we might be able to find a $\iota$ independent solution, that is $v_{2}(t):=v_{2}(t,\iota)$ which satisfies 
\bn
 \partial_{t}v_{2}+\frac{1}{\kappa}v_{2}^{2} -\phi&=&0. 
\en 
This is a Riccati equation which has the following solution (see the proof of Proposition 1 in \cite{Car-Jiam-2016}),
\bn
 v_{2}(t) = \sqrt{\kappa \phi} \frac{1+\zeta e^{2\beta(T-t)}}{1-\zeta e^{2\beta(T-t)}}, 
\en
where 
\bn
\zeta = \frac{ \varrho +\sqrt{\kappa \phi}}{\varrho-\sqrt{\kappa \phi}}, \qquad  \beta = \sqrt{\frac{\phi}{\kappa }}.
\en
Let $E_{t,\iota}$ represent expectation conditioned on $I_{t}=\iota$.

Using $v_{2}$, we can find a Feynman-Kac representation to the solution of (\ref{v1}),  
\bn  
v_{1}(t,\iota)  &=&E_{t,\iota}\Big[\int_{t}^{T}e^{\frac{1}{\kappa}\int_{t}^{s}v_{2}(u)du} I_{s}ds\Big] \\ 
&=&\int_{t}^{T}e^{\frac{1}{\kappa}\int_{t}^{s}v_{2}(u)du} E_{t,\iota}[I_{s}]ds.
\en
Again by Feynman-Kac formula we derive a solution to (\ref{v0}), 
\bn  
v_{0}(t,\iota)  &=&E_{t,\iota}\Big[\frac{1}{4\kappa}\int_{t}^{T}v^{2}_{1}(s,I_{s})ds\Big] \\ 
&=&\frac{1}{4\kappa}\int_{t}^{T}E_{t,\iota}\big[v^{2}_{1}(s,I_{s})\big]ds.
\en
\paragraph{Proof of Proposition \ref{verification}}
Note that $V$ is a classical solution to \ref{hjb}. By standard arguments (see e.g. Theorem 3.5.2 in \cite{pham}), in order  to prove that $V$ in (\ref{opt-V}) is the value function of (\ref{v-cost}), it is enough to show that $r^{*}$ is admissible and that 
\be \label{v-cond}
|V(t,\iota,c,x,p)|\leq C(1+\iota^{2}+c^{2}+x^{2}+p^{2}), \quad \textrm{for all } t\geq 0, \ \iota,c,x,p\in \re. 
\ee
Clearly $\sup_{t\in [0,T]}|v_{2}(t)|<\infty$. From our conditions on $I$ we have 
\bn
E_{\iota}\big[|I_{t}|] \leq C(1+|\iota|), \quad \textrm{for all } \iota \in \re, \  0\leq t\leq T, 
\en 
then we will have 
\bn
|xv_{1}(t,\iota)|&\leq& C|x|(1+|\iota|) \\
&\leq & C(1+\iota^{2}+x^{2}), \quad \textrm{for all } t\geq 0, \ \iota,x\in \re,
\en 
\bn
|v_{0}(t,\iota)|&\leq&  C(1+\iota^{2}), \quad \textrm{for all } t\geq 0, \ \iota \in \re. 
\en
and (\ref{v-cond}) follows. 
To prove that $r^{*}$ is admissible it is enough to show that $\int_{0}^{T}|r^{*}_{t}|dt<\infty$. 
Since $v_{2}$ is bounded we notice that 
\bn
|r^{*}_{t}|&\leq& \frac{1}{2\kappa}\Big(2|v_{2}(t)||X_{t}|+\int_{t}^{T}e^{\frac{1}{\kappa}\int_{t}^{s}|v_{2}(u)|du} E_{t,\iota}[|I_{s}|]ds\Big) \\
&\leq &C_{1}|X_{t}|+C_{2}T(1+|\iota|) \\
&\leq &(C_{2}+C_{1})\big(x+T(1+|\iota|)\big)+C_{1}\int_{0}^{t}|r_{s}|ds, 
\en
where we used (\ref{inv-eq}) in the last inequality. From Gronwall inequality we have
\bn
|r^{*}_{t}| \leq (C_{2}+C_{1})(x+T(1+|\iota|)) e^{C_{1}T}, 
\en
hence $r^{*}$ is admissible. 
%Then we have 
%\bn  
%v_{}(t,x,\iota) &=&E_{(t,x,\iota)}\Big[\int_{t}^{T}(\mu \iota_{s} x_{s} -\phi x_{s}^{2})ds-Ax_{T}^{2}\Big] \\ 
%&=&\mu x E_{(t,x,\iota)}\Big[\int_{t}^{T}e^{-M(s-t)}\iota_{s} ds\Big] -\phi x^{2}\int_{t}^{T}e^{-2M(s-t)}ds -Ax^{2}e^{-2M(T-t)}\\ 
%&=&\mu x \iota e^{(M+\gamma^{I})t} \int_{t}^{T}e^{-(M+\gamma^{I})s}ds  -\frac{\phi x^{2}}{2M}\big(1-e^{-2M(T-t)}\big) -Ax^{2}e^{-2M(T-t)}\\ 
%&=&\frac{\mu x \iota}{M+\gamma^{I}}\big(1- e^{-(M+\gamma^{I})(T-t)}\big)  -\frac{\phi x^{2}}{2M}\big(1-e^{-2M(T-t)}\big) -Ax^{2}e^{-2M(T-t)}\\ 
%&=&\frac{\mu x \iota}{M+\gamma^{I}}\big(1- e^{-(M+\gamma^{I})(T-t)}\big)  -\frac{\phi x^{2}}{2M}\big(1-e^{-2M(T-t)}\big) -Ax^{2}e^{-2M(T-t)}\\ 
%\en
\qed

\section*{Acknowledgements}
We are very grateful to anonymous referees and to the editors for careful reading of the manuscript, and for a number of useful comments and suggestions that significantly improved this paper.

\bibliographystyle{plain}
\printindex

\appendix

\clearpage
\section{Tables and complementary statistics}

\subsection{Composition of market participants groups}
\label{sec:compap}

\begin{table}[!ht]
  \centering
  {\bf High Fequency Traders}\\\medskip
  \begin{tabular}{|rc|cc|}\hline
Name & NASADQ-OMX & Market & Prop. \\
& member code(s)& Maker & Trader\\\hline
    All Options International B.V. & AOI& &\\
    Hardcastle Trading AG & HCT& &\\
    IMC Trading B.V & IMC, IMA& Yes & \\
    KCG Europe Limited & KEM, GEL& Yes & \\
    MMX Trading B.V & MMX& & \\
    Nyenburgh Holding B.V. & NYE& &\\
    Optiver VOF & OPV& & Yes\\
    Spire Europe Limited & SRE, SREA, SREB& & Yes\\
    SSW-Trading GmbH & IAT& & \\
    WEBB Traders B.V & WEB& &\\
    Wolverine Trading UK Ltd & WLV& &\\ \hline
  \end{tabular}
  \caption{Composition of the group of HFT used for empirical examples, and the composition of our ``high frequency market maker'' and ``high frequency proprietary traders'' subgroups.}
  \label{tab:compagents:HFT}
\end{table}
\begin{table}[!ht]
  \centering
  {\bf Global Investment Banks}\\ \medskip
  \begin{tabular}{|rc|}\hline
Name & NASADQ-OMX \\
& member code(s)\\\hline
 Barclays Capital Securities Limited Plc & BRC \\
    Citigroup Global Markets Limited & SAB \\
    Commerzbank AG & CBK \\
    Deutsche Bank AG & DBL \\
    HSBC Bank plc & HBC \\
    Merrill Lynch International & MLI \\
    Nomura International plc & NIP \\\hline
  \end{tabular}
  \caption{Composition of the group of Global Investment Banks used for empirical examples.}
  \label{tab:compagents:GIB}
\end{table}
\begin{table}[!ht]
  \centering
  {\bf Institutional Brokers}\\ \medskip
  \begin{tabular}{|rc|}\hline
Name & NASADQ-OMX\\
& member code(s)\\\hline
 ABG Sundal Collier ASA & ABC \\
              Citadel Securities (Europe) Limited & CDG \\
              Erik Penser Bankaktiebolag & EPB \\
              Jefferies International Limited & JEF \\
              Neonet Securities AB & NEO \\
              Remium Nordic AB & REM \\
              Timber Hill Europe AG & TMB \\\hline
  \end{tabular}
  \caption{Composition of the group of Institutional Brokers used for empirical examples.}
  \label{tab:compagents:brok}
\end{table}

\clearpage
\subsection{Complementary Statistics}
\label{sec:compstats}

\begin{table}[ht!]
  \centering
\begin{tabular}{lrrrrr}
\toprule
%{} & $\hat\gamma$ & $\hat\gamma$ & $\hat\gamma$ & $\hat\gamma$ & $\hat\gamma$ \\
{$\hat\gamma$ estimates} & 3 trades & 5 trades & 7 trades & 10 trades & 100 trades \\
\midrule
VOLVb.ST &        0.97 &        0.94 &        0.93 &         0.93 &          0.99 \\
NDA.ST   &        0.97 &        0.94 &        0.93 &         0.93 &          0.99 \\
ERICb.ST &        0.98 &        0.95 &        0.93 &         0.93 &          0.99 \\
HMb.ST   &        0.95 &        0.92 &        0.92 &         0.92 &          0.99 \\
ATCOa.ST &        0.95 &        0.92 &        0.91 &         0.91 &          0.99 \\
SWEDa.ST &        0.96 &        0.93 &        0.92 &         0.92 &          0.99 \\
SAND.ST  &        0.94 &        0.91 &        0.91 &         0.92 &          0.99 \\
SKFb.ST  &        0.92 &        0.90 &        0.90 &         0.91 &          0.99 \\
SEBa.ST  &        0.97 &        0.95 &        0.93 &         0.93 &          0.99 \\
NOKI.ST  &        0.73 &        0.82 &        0.86 &         0.90 &          0.99 \\
TLSN.ST  &        0.81 &        0.84 &        0.88 &         0.91 &          0.99 \\
ABB.ST   &        0.95 &        0.92 &        0.90 &         0.91 &          0.99 \\
AZN.ST   &        0.82 &        0.84 &        0.87 &         0.90 &          0.99 \\
\bottomrule
\end{tabular}
  \caption{Estimates of the speed of mean reversion $\gamma$, using different time scales.}
  \label{tab:imb:gamma}
\end{table}

\begin{table}[ht!]
  \centering\hspace*{-1em}
\begin{tabular}{lrr|rrrrr}
\toprule
{} & $\hat\kappa$ over   & $\hat{dt}$ (sec) & $\hat\sigma$ & $\hat\sigma$ & $\hat\sigma$ & $\hat\sigma$ & $\hat\sigma$ \\
{} & spread &                  & 3 trades & 5 trades & 7 trades & 10 trades & 100 trades \\
\midrule
VOLVb.ST &              0.088 &         5.30 &                  0.25 &                  0.23 &                  0.22 &                   0.19 &                    0.06 \\
NDA.ST   &              0.098 &         7.20 &                  0.26 &                  0.24 &                  0.22 &                   0.20 &                    0.07 \\
ERICb.ST &              0.092 &         6.60 &                  0.25 &                  0.23 &                  0.22 &                   0.19 &                    0.06 \\
HMb.ST   &              0.095 &         6.68 &                  0.27 &                  0.25 &                  0.22 &                   0.19 &                    0.06 \\
ATCOa.ST &              0.109 &         7.77 &                  0.27 &                  0.25 &                  0.23 &                   0.20 &                    0.06 \\
SWEDa.ST &              0.105 &         7.73 &                  0.27 &                  0.25 &                  0.23 &                   0.20 &                    0.06 \\
SAND.ST  &              0.101 &         7.24 &                  0.28 &                  0.25 &                  0.22 &                   0.19 &                    0.06 \\
SKFb.ST  &              0.108 &         8.53 &                  0.28 &                  0.25 &                  0.23 &                   0.20 &                    0.06 \\
SEBa.ST  &              0.099 &         9.13 &                  0.26 &                  0.24 &                  0.22 &                   0.19 &                    0.06 \\
NOKI.ST  &              0.172 &        10.24 &                  0.33 &                  0.26 &                  0.22 &                   0.19 &                    0.06 \\
TLSN.ST  &              0.134 &         7.74 &                  0.31 &                  0.26 &                  0.22 &                   0.19 &                    0.06 \\
ABB.ST   &              0.113 &        15.13 &                  0.28 &                  0.26 &                  0.24 &                   0.20 &                    0.07 \\
AZN.ST   &              0.163 &        15.51 &                  0.32 &                  0.26 &                  0.23 &                   0.19 &                    0.06 \\
\bottomrule
\end{tabular}  
  \caption{Estimate of $\kappa$ divided by the average bid-ask spread, average time between two trades, and level of noise in the estimated dynamics of liquidity. Each columns of level of noise $\sigma$ is estimated from a different time scale (i.e. number of trades). The decay in the estimates of $\sigma$ shows that innovations of the imbalance are sub-diffusive.}
  \label{tab:imb:sigma}
\end{table}

%\noindent Eyal Neuman: Department of Mathematics, University of Rochester, Rochester, 14627 NY, USA. eneuman4@ur.rochester.edu

%\medskip
%\noindent 
\end{document}